\documentclass[aps,prb,twocolumn,superscriptaddress,usenames,dvipsnames,longbibliography,floatfix,10pt]{revtex4-2}
\usepackage{hyperref}
\usepackage{bm}
\usepackage[T1]{fontenc}
\usepackage{amsmath}
\usepackage{amssymb}
\usepackage{graphicx}
\usepackage{subcaption}
\usepackage[dvipsnames]{xcolor}
\usepackage{tabularx}
\usepackage{multirow}
\usepackage{colortbl}
\usepackage{soul}
\usepackage{cancel}
\usepackage{textcomp}
\usepackage{gensymb}
\usepackage{siunitx}
\usepackage{comment}
\usepackage{tikz}
\usetikzlibrary{arrows.meta, positioning}

\usepackage[normalem]{ulem}
\hypersetup{colorlinks=true, linkcolor=blue, citecolor=red, urlcolor=magenta, pdftitle={Chirality from rutile crystals}, pdfauthor={S. Mamoudou Taganga, E. Bousquet}}\usepackage{natbib}
\definecolor{gray}{rgb}{0.8,0.8,0.8}

\begin{document}

\title{Chirality from rutile crystals}

\author{Samsia Mamoudou Taganga}
\affiliation{Physique Th\'eorique des Mat\'eriaux, Q-MAT, Universit\'e de Li\`ege, B-4000 Sart-Tilman, Belgium}

\author{Jean Milengezi}
\affiliation{Laboratoire de Physique des Solides et des Interfaces, Institut Sup\'e{}rieur P\'e{}dagogique de Bukavu, Democratic Republic of the Congo}

\author{Safari Amisi}
\affiliation{Physique Th\'eorique des Mat\'eriaux, Q-MAT, Universit\'e de Li\`ege, B-4000 Sart-Tilman, Belgium}
\affiliation{Laboratoire de Physique des Solides et des Interfaces, Institut Sup\'e{}rieur P\'e{}dagogique de Bukavu, Democratic Republic of the Congo}

\author{Eric Bousquet}
\affiliation{Physique Th\'eorique des Mat\'eriaux, Q-MAT, Universit\'e de Li\`ege, B-4000 Sart-Tilman, Belgium}

\begin{abstract}
We study from first-principles calculations the possible connectivity between the reported enantiomorphic $P4_12_12$ and $P4_32_12$ chiral phases and the hypothetical achiral reference rutile phase of the SnF$_2$ and TeO$_2$ crystals.
We show that the chiral phase can be explained through an unstable phonon of the rutile phase from the Z zone boundary point.
Several other instabilities are identified and a phase diagram of the different low symmetry phases that can be reached is built.
We then analyze whether the presence of a lone pair is an important key point to stabilize the chiral phases in the rutile phase by comparing the series SnO$_2$, GeF$_2$, GeO$_2$, PbCl$_2$, PbF$_2$, PbO$_2$ and SeO$_2$.
We found that the lone pair is a necessary but not sufficient parameter to stabilize the chiral phase and that there is a competition between the electronic and the elastic stability that emerges from a pseudo Jahn-Teller effect. 
\end{abstract}

\maketitle

\section{Introduction}

Structural chirality has attracted a renewed attention over the last years, in particular in relation with chiral phonons~\cite{Bousquet2025chirality,Juraschek2025chiralphonon}, the chiral induced spin selectivity~\cite{Ray1999CISS}, magnetic and polar skyrmions~\cite{Muhlbauer2009skyrmion}, Kramers-Weyl fermions~\cite{Bradlyn2016weylfermions,Hasan2021weylfermions} or altermagnetism~\cite{Smejkal2022altermagnetism,Smejkal2020halleffect}. 
In most of these cases, however, the handedness is fixed by the crystallization process and cannot be considered as an active degree of freedom of the solid~\cite{Flack2003chiralstructures}.
This is no longer the case if the chiral phase can be reached from an achiral parent phase through a displacive (or order-disorder) structural phase transition~\cite{Bousquet2025chirality, Fava2025k3nio2}. 
When a group-subgroup relationship exists between the two phases, the transition preserves the connectivity of the structure and the amplitude of the symmetry-lowering distortion can be taken as an order parameter, with a double-well energy landscape connecting the two handedness~\cite{Bousquet2025chirality, Fava2025k3nio2}. 
Such transitions are, however, scarce as the list of compounds for which an achiral-to-chiral phase transition toward an enantiomorphic pair that has actually been reported is limited to a few cases such as MgTi$_2$O$_4$, CsCuCl$_3$, Na$_2$Ca$_2$Si$_3$O$_9$, Ag$_4$P$_2$O$_7$ or K$_3$NiO$_2$~\cite{Bousquet2025chirality,GomezOrtiz2025pathways, Fava2025k3nio2, Fava2025handedness}. 
Hence, one can ask whether this scarcity has a physical or chemical origin, or whether these transitions have simply been overlooked~\cite{Bousquet2025chirality}.

A microscopic mechanism for such a transition has been proposed  recently by Fava {\it et al.} in the case of K$_3$NiO$_2$~\cite{Fava2025k3nio2}.
They have shown that a doubly degenerate unstable phonon mode at the $Z$ zone-boundary point of the achiral $P4_2/mnm$ parent phase drives the system towards the $P4_12_12$ or the $P4_32_12$ enantiomorphic phase, with four energetically equivalent wells in the energy surface. 
The degeneracy of the instability guarantees the existence of low-energy transition states connecting the two handedness, through which a circularly polarized electromagnetic field can select and switch the chirality of the crystal~\cite{Fava2025handedness}. 
Building on this picture, a screening strategy combining a pseudo-symmetry search with first-principles calculations has been proposed to identify systematically which chiral crystals may be connected to an achiral parent by a soft phonon mode~\cite{GomezOrtiz2025pathways}. 
Applied to a first set of compounds, it identified TeO$_2$, Sr$_2$As$_2$O$_7$ and Na$_2$SeO$_9$ as possible candidates, TeO$_2$ being predicted to be connected to a hypothetical rutile $P4_2/mnm$ parent phase, i.e. similarly to K$_3$NiO$_2$.

Whether chirality can be regarded as a new ferroic order parameter has been debated in this context~\cite{Bousquet2025chirality, spaldin2026, Juraschek2025chiralphonon} and it has been recently shown, by group theory and a systematic group-subgroup analysis, that an achiral phase cannot generate both members of an enantiomorphic pair through a zone-centre instability~\cite{GomezOrtiz2026ferroic}. 
Since a primary ferroic order parameter must transform according to a $\Gamma$-point irreducible representation, chirality cannot be a primary ferroic order when the chiral phase belongs to one of the 22 enantiomorphic space groups. 
The corollary is that the relevant instabilities necessarily occur at a finite wave vector, and hence that the search for new candidates has to be directed toward zone-boundary soft modes. 
Identifying which structural family and which chemical ingredients favour such instabilities is the challenge.

In the present work we address this question for two compounds, SnF$_2$ and TeO$_2$, whose reported chiral phases belong to the $P4_12_12/P4_32_12$ (for right- or left-handed crystals) enantiomorphic pair and both can come from a hypothetical rutile $P4_2/mnm$ parent phase~\cite{Baur01012007rutile}. 
Tin(II) fluoride is of particular interest since its $\gamma$ phase is chiral and is experimentally reached from the monoclinic $\alpha$ phase through a reconstructive transition~\cite{Denes1979alpha,Denes1980structure,Denes1980transitions,DENES198120thermalexpansion,Denes1981kinetics}, so that no displacive path to chirality is known for this compound. 
Paratellurite $\alpha$-TeO$_2$ has the same chiral symmetry~\cite{pathomas1988} and its rutile form has been stabilized epitaxially on a FeTe substrate~\cite{Peng2022teo2}, which makes the hypothetical achiral parent phase more than a computer design. 

In this work, we compute from first-principles density functional theory (DFT) calculations the phonon dispersion curves of the rutile phase of both compounds, condense and relax every unstable mode in order to build the hierarchy of the low-symmetry phases that can be reached, and show that the chiral phases originate from a doubly degenerate $Z_4$ instability. 
Finally, since the two compounds share a cation carrying a $ns^2$ lone pair, we extend the study to the {\it AX$_2$} series SnCl$_2$, SnO$_2$, GeF$_2$, GeCl$_2$, GeO$_2$, PbCl$_2$, PbF$_2$, PbO$_2$ and SeO$_2$ to determine whether the lone pair is the ingredient responsible for the chiral instability, and we analyze the result in terms of a pseudo Jahn-Teller effect competing with the elastic rigidity of the lattice. 
Beyond the two compounds studied here, the aim is to provide a chemical criterion that can be used to enlarge the still very short list of candidates for displacive achiral-to-chiral phase transitions.

%%%%%%%%%%%%%%%%%%%%%%%%%%%%%%%%%%%%%%%%%%%%%
\section{Technical details}
First-principles calculations were carried out using the ABINIT DFT  package~\cite{abinit1,abinit2,abinit-web,marques}, a planewave pseudopotential approach using the optimized norm-conserving Vanderbilt pseudopotentials (ONCVPSP) code~\cite{marques,hamann,Setten}, available on  pseudo-dojo server~\cite{pseudojo}. All of the present ab initio calculations were performed using our calculated theoretical lattice constants. The Perdew-Burke-Ernzerhof  revised for solids functional (GGA$-$PBEsol)~\cite{perdew} was employed as the method for modeling the exchange-correlation energy.

When condensing the unstable modes, we considered either for {\it Amm$2$} (38), {\it Pmn$2_1$} (31), {\it P$\Bar{4}2_1$m} (113), a 6-atom cell corresponding to $a_0$, $a_0$, $b_0$, with a sampling of 4$\times$4$\times$6  {\it k}-points. 
For the {\it Cmcm} (63), {\it P$4_12_12$} (92), {\it P$2_12_12_1$} (19), and {\it P$4_32_12$} (96), a 12-atom cell is considered corresponding to $a_0$, $a_0$, 2$b_0$, with a sampling of 4$\times$4$\times$3  {\it k}-points. 
For the {\it P$2/c$}  (13), and  {\it Pna$2_1$} (33),  a 12-atom cell is considered corresponding to 2$a_0$, $a_0$, $b_0$, with a sampling of 2$\times$4$\times$6  {\it k}-points. 
For the {\it C$2/$m} (12) and  {\it P$2_1$/c} (14), we considered a 24-atom cell corresponding to 2$a_0$, $a_0$, 2$b_0$, with a sampling of 2$\times$4$\times$3  {\it k}-points. 
For the {\it Pmnm}  (59), {\it P$4_2/m$} (84) and  {\it Pmm$2$} (51), a 24-atom cell is considered corresponding to 2$a_0$, 2$a_0$, $b_0$, with a sampling of 2$\times$2$\times$6  {\it k}-points. 
For the {\it I$\bar{4}2$d} (122), {\it Imma} (74) structures, a 48-atom supercell corresponding to 2$a_0$, 2$a_0$, 2$b_0$ with  a sampling of 2$\times$2$\times$3 {\it k}-points. 
 We explicitly checked that the relative energy of the different phases is well converged and independent of the choice of the supercell. Structural relaxations were performed until the forces and stresses were smaller than 10$^{-7}$ hartrees/bohr and  10$^{-7}$ hartrees/bohr$^{3}$ respectively. The phonon frequencies were calculated according to density functional perturbation theory (DFPT)~\cite{gonze}. The space group symmetry were checked by FINDSYM program~\cite{findsym} while to analyse  the group theory and relative contributions of different phonon modes to the distortions we employed ISOTROPY and AMPLIMODES codes, of Crystallographic Bilbao server~\cite{isotropycode,amplimodecode,isodistortcode}, used to perform a Symmetry-Adapted Mode (SAM) decomposition to analyze structural distortions in crystal lattices.

%%%%%%%%%%%%%%%%%%%%%%%%%%%%%%%%%%%%%%

\section{\texorpdfstring{$\bm{\mathrm{SnF}}_{2}$}{SnF2}}
Tin(II) fluoride exhibits three polymorph forms, whose structures and phase transitions have been characterized in detail by Den\`es and co-workers~\cite{Denes1979alpha,Denes1980structure,Denes1980transitions,DENES198120thermalexpansion,Denes1981kinetics}. 
The room temperature structure, $\alpha$-SnF$_2$, is monoclinic with the centrosymmetric, and hence achiral, $C2/c$ (15) space group.
On heating, it transforms at $428$~K into the tetragonal $\gamma$ phase, which belongs to one of the two enantiomorphic space groups $P4_12_12$ (92) or $P4_32_12$ (96) and remains stable up to melting at $486$~K.
On cooling from the melt, SnF$_2$ crystallizes directly in the $\gamma$ phase at $458$~K, which then survives as a metastable form down to $339$~K where it converts reversibly into the orthorhombic $\beta$ phase of {\it P$2_12_12_1$} (19) symmetry.
Since {\it P$2_12_12_1$} is a subgroup of both $P4_12_12$ and $P4_32_12$ enantiomorphic space groups, this last transformation is displacive, $\beta$-SnF$_2$ being an orthorhombic deformation of the chiral $\gamma$ phase. 
We note that $\beta$-SnF$_2$ is itself chiral, but that {\it P$2_12_12_1$} is a non-enantiomorphic Sohncke group, such that its handedness is not fixed by its symmetry \cite{Fecher2022chirality}. 
The $\alpha \rightarrow \gamma$ transition, on the contrary, is reconstructive~\cite{Denes1980transitions,Denes1981kinetics} as it breaks and reforms first-neighbor bonds and as no group-subgroup relation exists between $C2/c$ and $P4_12_12/P4_32_12$.
The route through which SnF$_2$ becomes chiral is therefore not a displacive one, and no achiral parent phase from which the enantiomorphic phases could develop continuously has been reported so far.

Following the mechanism identified in K$_3$NiO$_2$~\cite{Fava2025k3nio2}, we examine whether such a parent phase can nevertheless be identified for SnF$_2$, and whether the $P4_12_12/P4_32_12$ phases can be described as arising from a phonon-driven displacive transition starting from the rutile $P4_2/mnm$ (136) structure. 
This rutile phase of SnF$_2$ has not been observed experimentally and is used here as a hypothetical high-symmetry reference.

\begin{figure*}[htbp]
    \centering
    \begin{subfigure}[t]{0.27\linewidth}
        \begin{subfigure}[b]{0.48\linewidth}
            \caption*{(A)}
            \centering
            \includegraphics[width=0.95\linewidth]{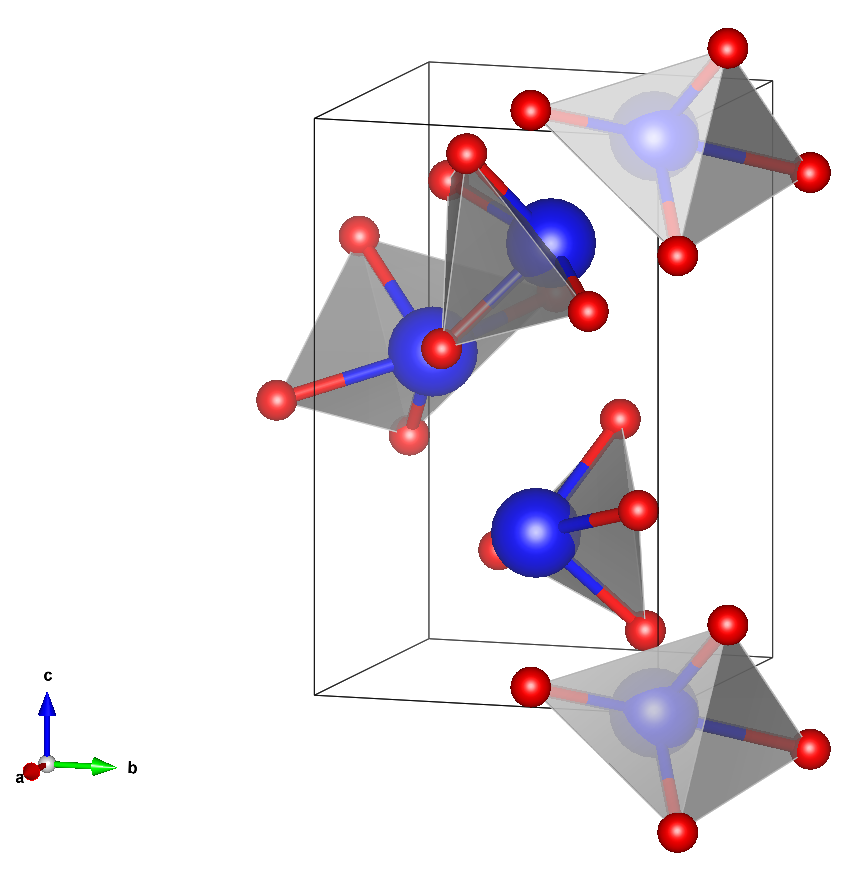}
        \end{subfigure}
        \begin{subfigure}[b]{0.48\linewidth}
            \centering
            \caption*{(B)}
            \includegraphics[width=0.95\linewidth]{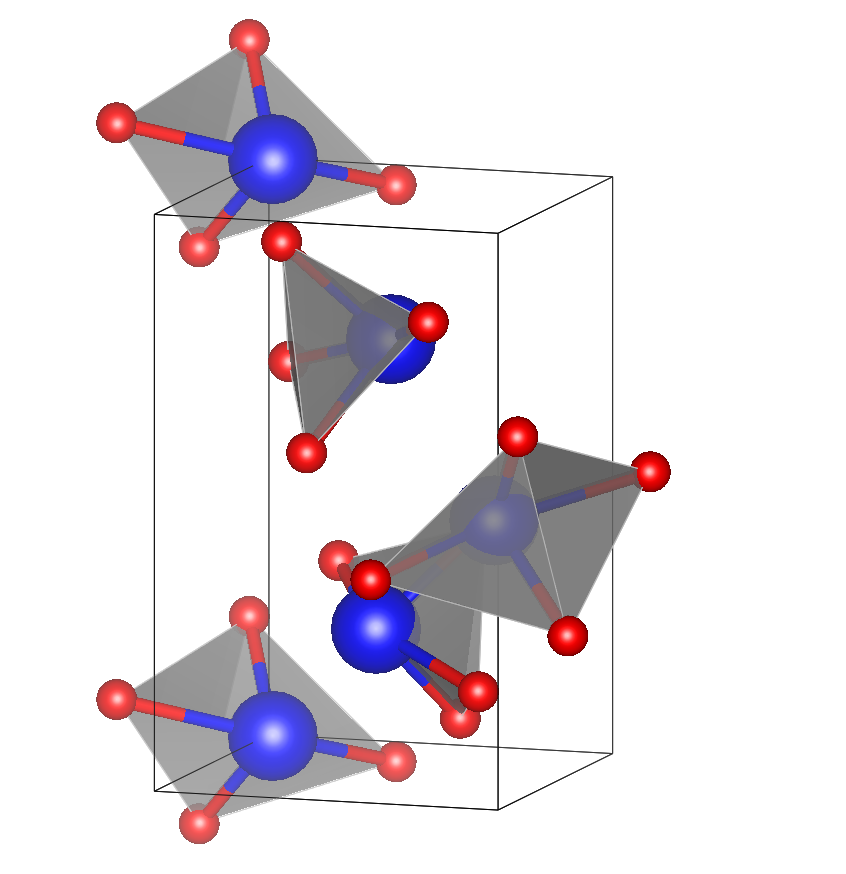}
        \end{subfigure}
    \end{subfigure}
        \hfill
    \begin{subfigure}[b]{0.35\linewidth}
        \caption{}
        \centering
        \includegraphics[width=0.9\linewidth]{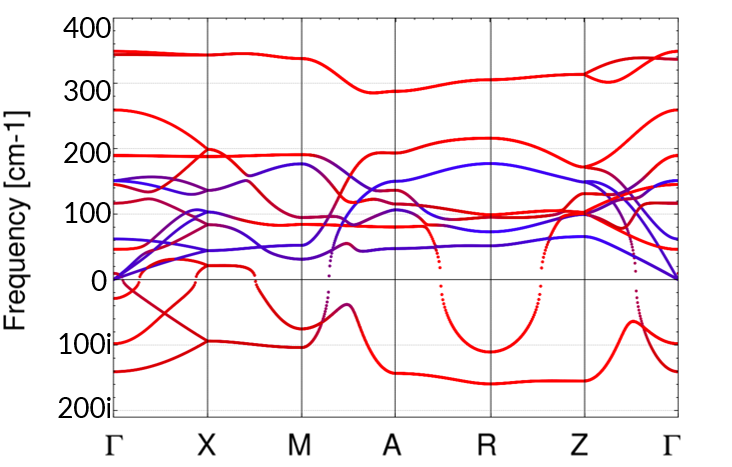}
        \label{fig:snf2_phonon}
    \end{subfigure}
    \hfill
    \begin{subfigure}[b]{0.35\linewidth}
        \caption{}
        \centering
        \includegraphics[width=0.95\linewidth]{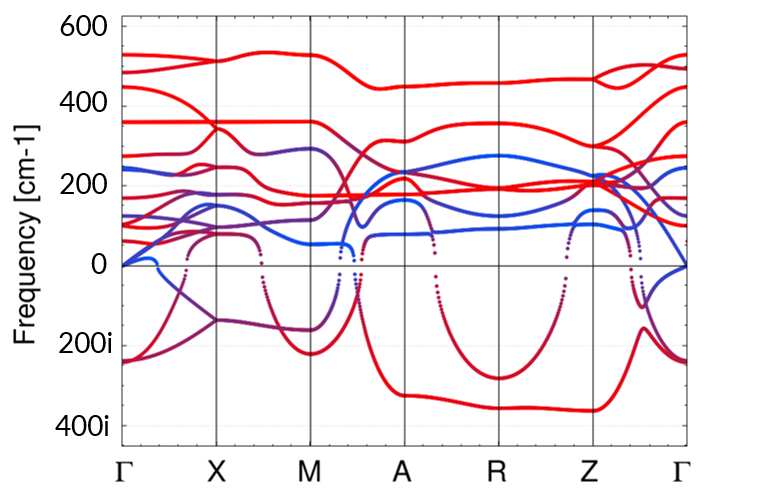}
        \label{fig:TeO2_phonon}
    \end{subfigure}

    \begin{subfigure}[b]{0.27\linewidth}
        \begin{subfigure}[b]{0.48\linewidth}
            \centering
            \caption*{(C)}
            \includegraphics[width=\linewidth]{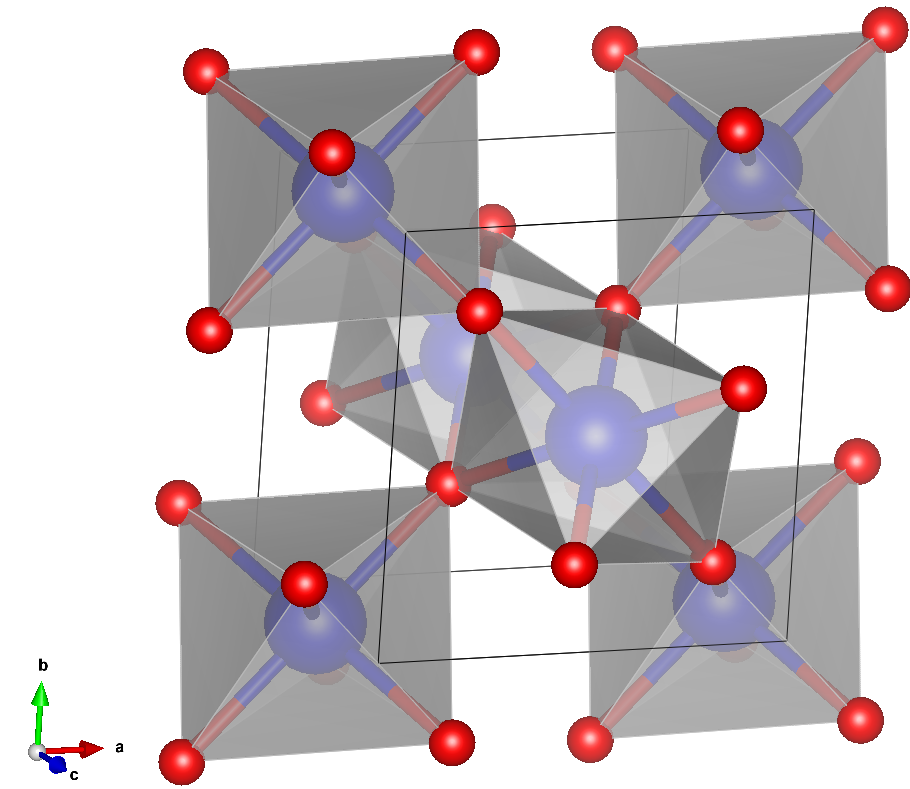}
        \end{subfigure}
        \hfill
        \begin{subfigure}[b]{0.48\linewidth}
            \centering
            \caption*{(D)}
            \includegraphics[width=0.9\linewidth]{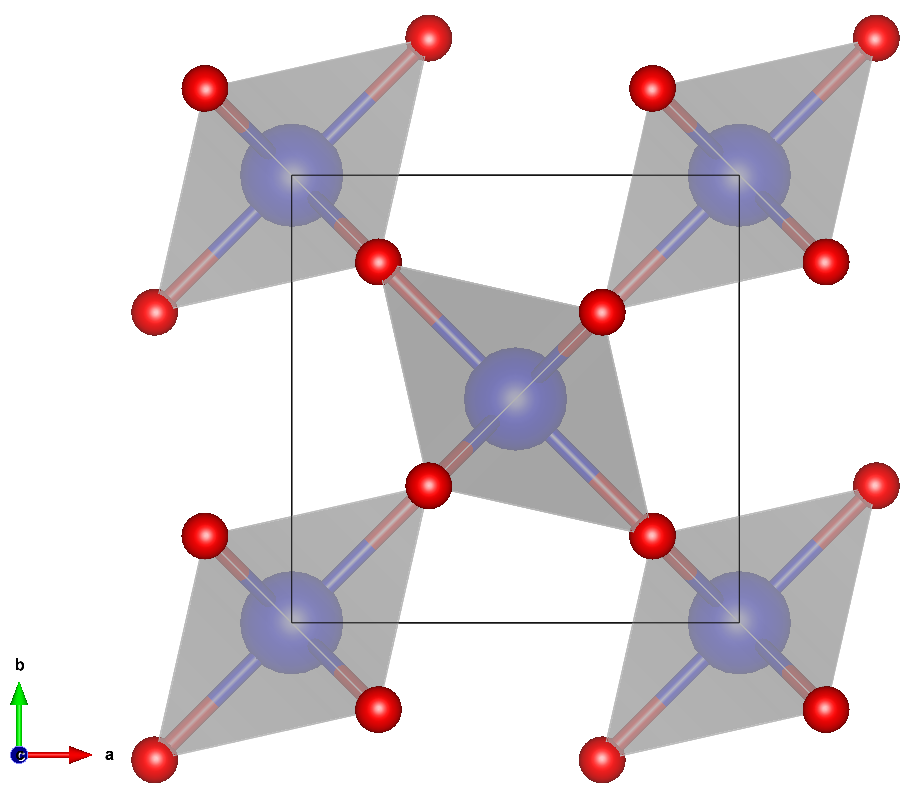}
        \end{subfigure}
    \end{subfigure}
    \hfill
    \begin{subfigure}[b]{0.35\linewidth}
        \caption{}
        \centering
        \includegraphics[width=0.9\linewidth]{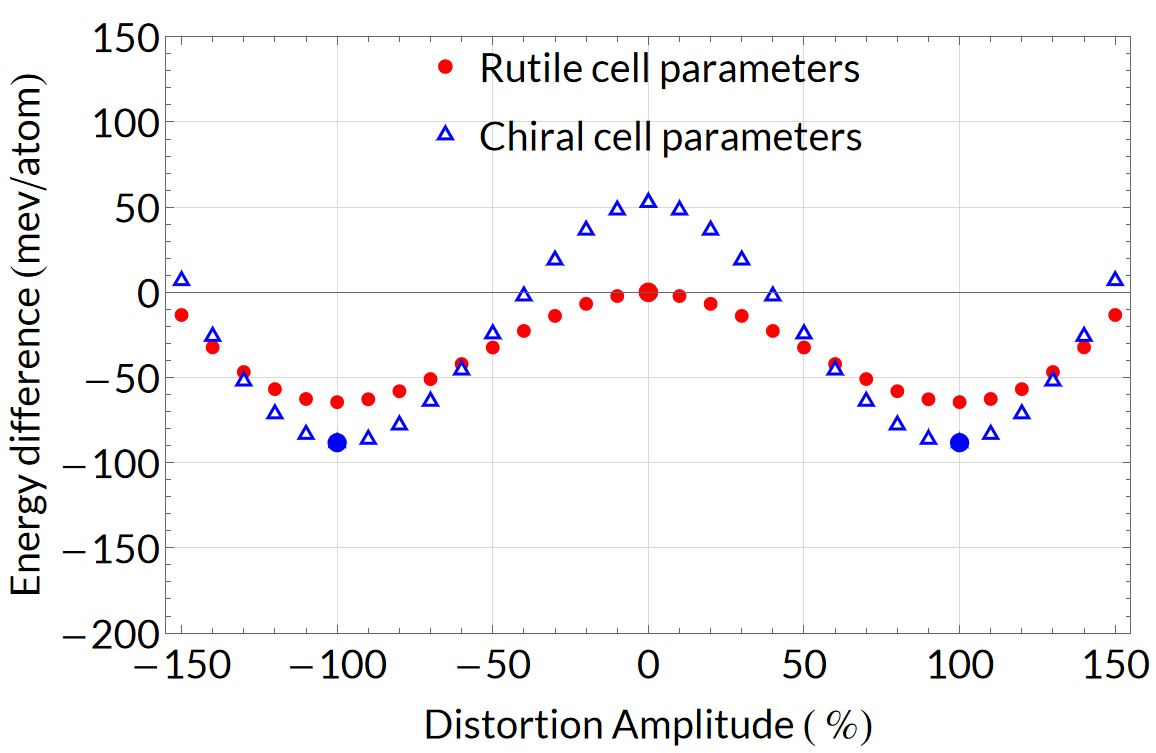}
        \label{fig:snf2_energydiff_phase}
    \end{subfigure}
    \begin{subfigure}[b]{0.35\linewidth}
        \caption{}
        \centering
        \includegraphics[width=0.9\linewidth]{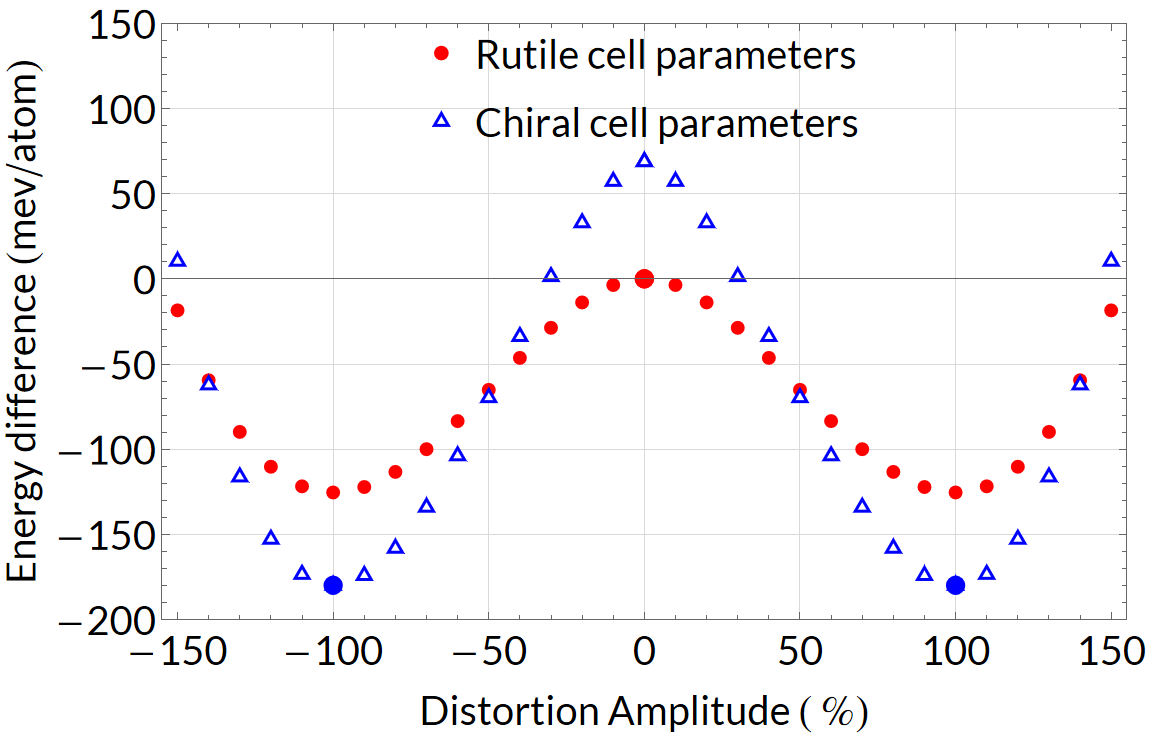}
        \label{fig:teo2_energydiff_phase}
    \end{subfigure}
    
    \caption{Structure  of $\gamma-$SnF\(_2\) in chiral space groups (A) $P4_12_12$ (92) and (B) $P4_32_12$ (96). Structure of hypothetical rutile SnF\(_2\) in space group $P4_2/mnm$ (136) is represented in (C) standard view and (D) top view. Sn/Te and F/O atoms are represented in blue and red, respectively.
    Phonon band structure of rutile $(a)$ SnF$_2$ and $(b)$ TeO$_2$ along the path $\Gamma \;-\; X \;-\; M \;-\;A\;-\; R\;-\; Z\;-\; \Gamma$. The imaginary frequencies, corresponding to unstable phonon modes, are plotted as negative numbers. They correspond to structural instabilities, indicating that this rutile phase is highly unstable at 0K. 
    The figure $(c)$ and $(d)$ corresponds to the energy difference with the rutile phase, respectively in SnF\(_2\) and TeO\(_2\), depending on the distortion amplitude towards chiral phases. 100\% and -100\% are the relaxed atomic positions of the right and left enantiomers, respectively, for the considered cell parameters. The red plot corresponds to the energy plot with the rutile cell parameters, while the blue plot corresponds to the chiral cell parameters. The energy of the undistorted rutile phase is taken as the zero reference.}
    \label{fig:snf2_struct_chiral}
\end{figure*}

\subsection{Phonon dispersion of Rutile high symmetry reference structure}

We first relaxed SnF$_2$ within the rutile $P4_2/mnm$ symmetry. The resulting reference cell contains six atoms, i.e. two formula units, the Sn cations being octahedrally coordinated by six F anions with the SnF$_6$ octahedra sharing edges along the $c$ axis. Its phonon dispersion curves, computed within DFPT on a $\mathbf{q}$-point mesh along the $\Gamma - X - M - A - R - Z - \Gamma$ path, are reported in Fig.~\ref{fig:snf2_phonon}.
Among the 18 phonon branches, a large number are found to be imaginary. 
The rutile form of SnF$_2$ is therefore far from being a local minimum of the energy surface, which could be consistent with the fact that it has never been reported experimentally. 

The unstable modes frequencies and irrep are reported in Tab.~\ref{tab:phonon_instable_points}.
At the zone centre, two instabilities are present, a non-degenerate $\Gamma_2^-$ mode at $141i$~\si{\per\centi\meter} and a doubly degenerate $\Gamma_5^-$ mode at $98i$~\si{\per\centi\meter}. 
The $\Gamma_2^-$ mode ($B_{1u}$ symmetry) is a non-polar structural distortion. Although it breaks the inversion symmetry and leads to the non-centrosymmetric $P\bar{4}2_1m$ (113), it does not generate a macroscopic polarization. It can be described as an internal stretching of SnF$_6$ units, where local dipoles cancel each other out in the unit cell.
The $\Gamma_5^-$ mode is the in-plane polar mode of the rutile structure and its condensation along the $(0,a)$ direction gives the polar {\it Pmn$2_1$} (31) phase, i.e. a ferroelectric-like instability of the type already discussed in other rutile compounds. Its condensation along $(a,a)$ direction give the {\it Amm$2$} (38) phase and along $(a,b)$, $Pm$ (6).

At the zone boundary, the doubly degenerate $X_1$ mode at $94i$~\si{\per\centi\meter} correspond to antipolar and antiferrodistortive motions that double the cell in the $(a,b)$ plane. 
It is dominated by scissoring of the Sn$-$F bonds. 
Due to its degeneracy, this mode can drive the system towards different structural phases depending on the condensation direction. 
A condensation along the $(a,0;0,0)$ direction leads to the $Pna2_1$ (33) phase and along $(a,a;0,0)$, the centrosymmetric $P2/c$ (13) phase.
The two doubly degenerate modes observed at $M$ are $M_2^- M_3^-$ at $103i$~\si{\per\centi\meter} and $M_5^-$ at $75i$~\si{\per\centi\meter} correspond to antiferrodistortive motions that quadruple the cell in the $(a,b)$ plane. 
The $A$ point hosts a doubly degenerate $A_4$ unstable mode at $143i$~\si{\per\centi\meter}. 
It corresponds to a complex zone-corner distortion that doubles the cell periodicity in all three directions. 
This mode combines octahedral tilting and cation-anion translations.
The strongest instability of the whole spectrum is found at $R$, where two doubly degenerate modes of $R_1^+$ symmetry are obtained at $159i$ and $110i$~\si{\per\centi\meter}. 
It is characterized by an anion-driven tilting of the SnF$_6$ octahedra where the Sn cations remain stationary on their inversion sites.

The mode of interest for the present work is the doubly degenerate $Z_4$ instability found at $154i$~\si{\per\centi\meter} at the $Z = (0,0,\frac{1}{2})$ zone-boundary point. 
It has the same symmetry, at the same point of the Brillouin zone and from the same $P4_2/mnm$ space group (but different structure), as the mode driving the achiral-to-chiral transition of K$_3$NiO$_2$~\cite{Fava2025k3nio2} and as the one predicted for TeO$_2$~\cite{GomezOrtiz2025pathways}. 
It is the only instability of the spectrum that can lead to the $P4_12_12/P4_32_12$ enantiomorphic pair, and its double degeneracy is what allows both handedness to be generated from a single unstable mode. 

The amplitude of the imaginary frequencies does not, however, give the order of the related energies when the modes are condensed as the harmonic frequency alone does not tell us about the anharmonic energy gain nor about the strain relaxation that accompanies the condensation. 
Each instability has therefore to be condensed and relaxed separately, which is the object of the next section.

\subsection{Phonon condensation and phase diagram}

\begin{table*}[htbp]
    \centering
    \renewcommand{\arraystretch}{1.5}
   \begin{tabular*}{\textwidth}{@{\extracolsep{\fill}}l c c c c c c c c}
        \hline\hline
        \textbf{$\mathbf{k}-$Point} & \multicolumn{2}{c}{$\mathbf{\omega}$} & \textbf{Main modes} & \textbf{Domain} & \textbf{Space Group} & \multicolumn{2}{c}{\textbf{$\Delta E$}} & \textbf{Additional modes} \\
        & \multicolumn{2}{c}{(1/cm)} & & & &  \multicolumn{2}{c}{(meV/atom)} & \\
         & {\footnotesize SnF\(_2\)} & {\footnotesize TeO\(_2\)} & & & & {\footnotesize SnF\(_2\)} & {\footnotesize TeO\(_2\)} & \\
        \hline
        $\Gamma$: 2  & 141$i$ &237$i$& $\Gamma_2^-$ & $(a)$ & {\it P$\bar{4}2_1$m} (113)  & -67 & -150 & $\Gamma_1^+$\\
         & 98$i$* &241$i$* & $\Gamma_5^-$ &  $(a,0)$ & {\it Pmn$2_1$} (31)  &  -52 & -93 & $\Gamma_1^+, \,\Gamma_2^+$ \\
         & & & &$ (a,a)$ &  {\it A$mm2$} (38) & -16 & -39 &  $\Gamma_1^+, \,\Gamma_4^+$  \\
        $X$: 1  & 94$i$* & 135$i$* & $X_1$ & $(a,0;0,0)$ & {\it P$na2_1$} (33) &  -88 & -65 & $\Gamma_1^+, \,\Gamma_2^+, \, \Gamma_5^-$\\
        & & & &  $(a,a;0,0)$ & {\it P$2/c$} (13) &  -5& -8 & $\Gamma_1^+, \,\Gamma_2^+, \,\Gamma_3^+, \,\Gamma_4^+$\\
        %& & &  $(a,b;a,b)$ & {\it Aem$2$} (39) & -10 & \\
        %& & &  $(a,a;-a,a)$ & {\it P$4_2/n$} (56) & -63 & \\
        %& & & $(a,b;0,0)$ & {\it Pc} (7) &  -87 & $\Gamma_1^+, \,\Gamma_2^+$\\
        %
        $Z$: 1  & 154$i$* & 363$i$* & {\it Z$_4$} & $(a,0)$& {\it Cmcm} (63)  & -35 & -67 & $\Gamma_1^+, \,\Gamma_4^+$ \\
        & & & &$ (a,a)$ &  {\it P$4_1 2_1 2$} (92) & \textbf{-89} & \textbf{-180} & $\Gamma_1^+$\\
        & & & & $(a,-a)$ &   {\it P$4_3 2_1 2$} (96) & \textbf{-89} & \textbf{-180} & $\Gamma_1^+$\\
        $M$: 2 & 103$i$* & 161$i$* & $M_2^- M_3^-$ & $(a,b)$ & {\it Pmmn} (59)&  -53 & -131 & $\Gamma_1^+, \,\Gamma_4^+$\\
        &  75$i$* &220$i$* & $M_5^-$& $(a,0)$&  {\it P$4_2$/m} (84)& -48 & -69 & $\Gamma_1^+, \,\Gamma_3^+$\\
        & & & &$ (a,a)$ &  {\it Pmm$2$} (51) & -56 &-33 & $\Gamma_1^+, \,\Gamma_4^+$\\
        $R$: 2  & 159$i$* & 356$i$* & $R_1^+$ & $(a,0;0,0)$ &  {\it P$2_1$/c} (14) & -84 & -179 & $\Gamma_1^+, \,\Gamma_2^+, \, \Gamma_5^+$\\
        & &  & & $(a,a;0,0)$ & {\it C$2/m$} (12) & -49 & -56 & $\Gamma_1^+, \,\Gamma_2^+, \,\Gamma_3^+, \,\Gamma_4^+$ \\
        %& & & $(a,b;0,0)$ & {\it P$\bar{1}$} (2) & -69 & $\Gamma_1^+, \,\Gamma_2^+, \,\Gamma_3^+, \,\Gamma_4^+, \, \Gamma_5^+$ \\
        %
        $A$: 1  & 143$i$* & 325$i$* & $A_4$ &  $(a,0)$ &  {\it Imma} (74) &  -27 & -46 & $\Gamma_1^+, \,\Gamma_4^+$\\
        & &  &  &$(a,a)$ & {\it I$\bar{4}2$d} (122) & -62 &-128 & $\Gamma_1^+$ \\
        \hline
    \end{tabular*}
    \caption{Summary of unstable phonon frequencies (second column) at high-symmetry {\it k}-points (first column), their main domains (fourth column) and corresponding space group (fifth column) and the associated gain of energy after full DFT relaxation calculations (sixth column) for the hypothetical rutile phase of SnF\(_2\)  and TeO\(_2\).  
    Eigenfrequencies are given in \si{\per\centi\meter}, an asterisk indicates the mode is degenerated twice. 
    $\Delta E$ corresponds to the energy difference from the rutile phase to the lower symmetry phase, in \si{\milli\electronvolt}/atom. A negative \(\Delta E\) indicates a more stable distorted phase.
    }
    \label{tab:phonon_instable_points}
\end{table*}

\begin{figure}[h!]
\centering
\begin{tikzpicture}[
    x=0.7cm, %pour ajuter l'etirement et la compression
    y=-0.1cm, 
    gpbox/.style={draw, fill=white, inner sep=2pt, font=\scriptsize, minimum height=1.2em},
    redbox/.style={gpbox, draw=red!70!black},
    greenbox/.style={gpbox, draw=green!50!black},
    lbl/.style={blue!70!black, font=\scriptsize\itshape, node distance=0pt},
]

    % AXE VERTICAL AVEC COMPRESSION
    % Partie 1
    %\draw[thick] (0,-3) -- (0,3);
    
    % Brisure d'axe 
    %\draw[thick] (-0.2, 2) -- (0.2, 4);
    %\draw[thick] (-0.2, 4) -- (0.2, 6);
    
    % Partie basse remontée (on simule que 30 est juste après la brisure)
    % L'axe continue de "10" (qui représente 30) jusqu'à "75" (qui représente 95)
    \draw[thick, -{Stealth}] (0,-3) -- (0,100) node[below, align=left, font=\tiny] {Energy Gain\\(meV/atom)};
    
    % --- GRADUATIONS ET LIGNES (COORDONNÉES RECALCULÉES) ---
    % Le 0
    \draw[thick] (-0.1, 0) -- (0.1, 0);
    \node[left=2pt, font=\tiny\bfseries] at (0, 0) {0};
    \draw[thick] (0,0) -- (10,0); % Ligne du haut (Rutile)

    % Fonction de mapping pour remonter le bloc : 
    \foreach \val/\yplot in {10, 20, 30, 40, 50, 60, 70, 80, 90, 100} {
        \draw[thick] (-0.1, \yplot) -- (0.1, \yplot);
        \node[left=2pt, font=\tiny\bfseries] at (0, \yplot) {\val};
        \draw[dotted, black!40] (0.1, \yplot) -- (10, \yplot); 
    }

    % ELÉMENTS DU DIAGRAMME
    
    % Niveau 0
    \node[gpbox] at (5,0) {$P4_2/mnm$ (136)};

    % BLOC DU BAS 
    \node[gpbox] (p113) at (1.5,67) {$P\bar{4}2_1m$ (113)};
    \node[lbl, above=of p113] {$\Gamma_2^-(a)$};
    
    \node[gpbox] (p31) at (1.5,52) {$Pmn2_1$ (31)};
    \node[lbl, above=of p31] {$\Gamma_5^-(a,0)$};

    \node[gpbox] (p38) at (1.5,16) {$Amm2$ (38)};
    \node[lbl, above=of p38] {$\Gamma_5^-(a,a)$};
    
    \node[gpbox] (p13) at (9,5) {$P2/c$ (13)};
    \node[lbl, left=of p13] {$X_1(a,a;0,0)$};

    \node[gpbox] (p33) at (9,88) {$Pna2_1$ (33)};
    \node[lbl, below=of p33] {$X_1 (a,0;0,0)$};

    \node[gpbox] (c63) at (4.2,35) {$Cmcm$ (63)};
    \node[lbl, above=of c63] {$Z_4(a,0)$};
    
    \node[redbox] (p92) at (4.8,89) {$P4_{1/3}2_12$ (92/96)};
    \node[lbl, above=of p92] {$Z_4(a, \pm a)$};

     \node[gpbox] (p59) at (4.2,53) {$Pmmm$ (59)};
    \node[lbl, above=of p59] {$M_2^-\ M_3^-(a,b)$};

    \node[gpbox] (p84) at (6.65,43) {$P4_2/m$ (84)};
    \node[lbl, above=of p84] {$M_5^-(a,0)$};
    
    \node[gpbox] (p51) at (6.65,56) {$Pmm2$ (51)};
    \node[lbl, above=of p51] {$M_5^-(a,a)$};

    \node[gpbox] (p12) at (9,49) {$C2/m$ (12)};
    \node[lbl, above=of p12] {$R_1^+(a,a;0,0)$};

    \node[gpbox] (p14) at (9,84) {$P2_1/c$ (14)};
    \node[lbl, above=of p14] {$R_1^+(a,0;0,0)$};
    
    \node[gpbox] (i74) at (9,27) {$Imma$ (74)};
    \node[lbl, above=of i74] {$A_4(a,0)$};

    \node[gpbox] (i122) at (9,62) {$I\bar{4}2d$ (122)};
    \node[lbl, above=of i122] {$A_4(a,a)$};

    \node[greenbox] (p15) at (1.5,93) {$C2/c$ (15)};
    \node[lbl, left=of p15] {$\alpha$};

    \node[greenbox] (p19) at (1.6,89) {$P2_12_12_1$ (19)};
    \node[lbl, above=of p19] {$\beta$};

\end{tikzpicture}
\caption{Hierarchy of energy gains from the rutile parent phase in \si{\milli\electronvolt}/atom as a reference. For each phase, the space group and the main condensing phonon mode (in blue) are given. The red box highlights the chiral phases, associated with the largest energy gain. The green boxes highlight $\alpha-$ and $\beta-$ SnF$_2$ phases, observed experimentally. They respectively have an energy gain of -93
and -88 \si{\milli\electronvolt}/atom.}
\label{fig:snf2_phase_diagram}
\end{figure}
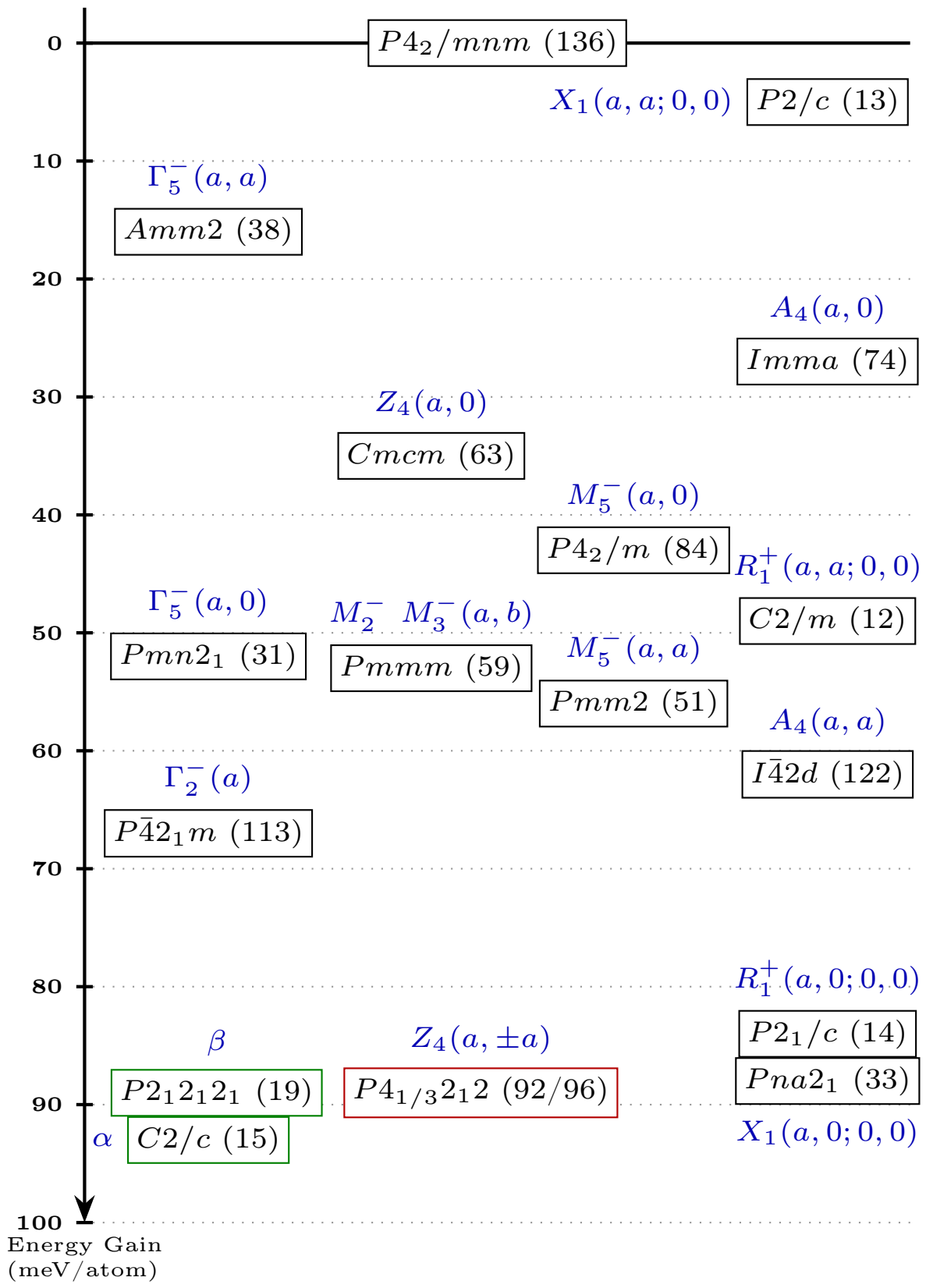

Each unstable mode of Tab.~\ref{tab:phonon_instable_points} was condensed into the rutile reference and the resulting structure was then fully relaxed, in both atomic positions and cell parameters.
For the degenerate instabilities, different order-parameter directions allowed by the irreducible representation, reported as the domain in Tab.~\ref{tab:phonon_instable_points}, were condensed separately since they lead to distinct isotropy subgroups. 
The resulting distortion with respect to the rutile reference was decomposed into SAM, which gives the secondary modes allowed to appear by symmetry even if with stable frequencies (last column of Tab.~\ref{tab:phonon_instable_points}). 
The resulting hierarchy of energies is summarized in Fig.~\ref{fig:snf2_phase_diagram}.

The main observation from these calculations is that the chiral phases are the most stable of the whole set. 
Condensing the doubly degenerate $Z_4$ mode along the diagonal $(a,a)$ and $(a,-a)$ directions gives the $P4_12_12$ (92) and $P4_32_12$ (96) chiral phases with an energy gain of $89$~\si{\milli\electronvolt}/atom, the largest of all the instabilities considered. 
The two values are strictly identical, as they must be since the two structures are enantiomorphs of each other, which also provides a  numerical check of the calculations. 
Hence, not only can the chiral phases of SnF$_2$ be connected to the rutile reference through a single unstable phonon mode, but they also correspond to the deepest minimum that can be reached from it.

The non-diagonal directions of the same $Z_4$ mode lead to a different phase. 
When only one component of the order parameter is condensed, i.e. along $(0,a)$, the system relaxes into the achiral orthorhombic $Cmcm$ (63) phase with an energy gain of $35$~\si{\milli\electronvolt}/atom. 
As in K$_3$NiO$_2$~\cite{Fava2025k3nio2,Fava2025handedness}, the role of this phase is not that of a competing ground state but that of a transition state, i.e. the achiral configuration through which the two handedness can be interconverted without going back to the much higher rutile reference.
The associated barrier is $54$~\si{\milli\electronvolt}/atom and is discussed below together with the energy wells.

The chiral phases are, however, not isolated in energy. 
The $X_1$ instability relaxes into the orthorhombic $Pna2_1$  phase lying only $1$~\si{\milli\electronvolt}/atom above the chiral phases.
The strongest instability of the phonon spectrum, the $R_1^+$ mode at $159i$~\si{\per\centi\meter}, gives a $P2_1/c$ (14) phase $5$~\si{\milli\electronvolt}/atom higher than the chiral ones.  
The remaining instabilities are clearly higher in energy, between $5$ and $67$~\si{\milli\electronvolt}/atom below the parent phase. 
In particular, the polar $Pmn2_1$ phase coming from the $\Gamma_5^-$ mode lies $37$~\si{\milli\electronvolt}/atom above the chiral phases, such that the rutile phase of SnF$_2$ has a marked preference for the chiral distortion over the ferroelectric one. This whole energy diagram lies above the two other phases observed experimentally. $\beta-$SnF\(_2\) belonging in the orthorhombic \(P2_12_12_1\) space group is a distortion of the tetragonal chiral phase. They are degenerate, explaining the reversible transition observed experimentally between $339$~K and $383$~K. $\alpha-$SnF\(_2\) is the experimental room temperature structure and exhibit the lowest energy, $5$~\si{\milli\electronvolt}/atom lower than the chiral phases.

The SAM decomposition gives us some extra details.
The fully symmetric (or invariant) $\Gamma_1^+$ mode appears in every case, which simply reflects the rutile cell deformation accompanying the condensation as well as the symmetric distortion of polyhedra. 
For the chiral phases it is the only secondary mode, i.e. the distortion from the rutile phase to $P4_12_12/P4_32_12$ is entirely described by the $Z_4$ order parameter together with the isotropic relaxation of the cell, with no additional symmetry-adapted distortions. 
This is a much simpler situation than for the $P2/c$ (13) and $C2/m$ (12) phases, which involve up to four secondary modes, and it is what allows the atomic distortion and the cell relaxation to be disentangled in the next section.

%%%%%%%%%%%%%%%%%%%%%%%%%%%%%%%%%%
\subsection{Energy wells of the chiral phonon mode}

The SAM analysis has shown that the distortion connecting the rutile reference to the chiral phases involves two main ingredients, the $Z_4$ order parameter and the fully symmetric relaxation of the cell. 
These two ingredients can be separated by computing the energy as a function of the $Z_4$ amplitude at a fixed cell, and by repeating the calculation for two different choices of that cell. 

In the first case, the lattice parameters are constrained to those of the optimized rutile parent phase doubled along the $z$ axis, $a = b = 5.471$~\AA{} and $c = 7.600$~\AA, and only the atoms are displaced along the $Z_4$ eigenvector. 
In the second case, the lattice parameters are those of the fully relaxed chiral structure, $a = b = 5.075$~\AA{} and $c = 8.188$~\AA . These values agree closely with the experimentally measured lattice parameters,
$a=b=5.073$~\AA{} and $c=8.491$~\AA \cite{Denes1980structure}, especially in $x$ and $y$ direction. The calculated $c$ parameter is about $3.6\%$ shorter than the experimental value, a difference consistent with the comparison between a 0 K DFT-relaxed structure and a finite-temperature ($T=353$ K) experiment. 
In both cases, the amplitude is normalized so that $\pm 100\%$ corresponds to the atomic positions of the two enantiomers for the cell under consideration, and $0\%$ to the rutile atomic positions.
The two cells differ considerably: going from the rutile to the chiral one corresponds to a contraction of $7.2\%$ in the $ab$ plane and an elongation of $7.7\%$ along $c$, i.e. a change of the $c/a$ ratio from $1.389$ to $1.613$ and a reduction of the volume by $7.3\%$. 
This allows us to see the influence of the $Z_4$ mode and of the cell relaxation during the achiral to chiral transition and we can see that the chiral phase is not a small perturbation of the rutile cell.

The energies obtained along those two cases are reported in Fig.~\ref{fig:snf2_energydiff_phase}.
Both curves show the double-well shape expected from an unstable phonon mode, with the two minima strictly symmetric with respect to the achiral $0\%$ configuration. 
The two well cases are, however, of very different nature. 
At the rutile cell (red curve), the $Z_4$ distortion is unstable but the energy gain is large and the minimum is reached for a displacement of the tin atoms of only $\pm 0.012$~\AA. 
At the chiral cell (blue curve), the double well is deeper and it is reached for tin displacements of $\pm 0.083$~\AA, i.e. seven times larger. 
In addition, the blue curve does not start from zero: holding the atoms at their rutile positions while imposing the chiral cell costs elastic energy, which is why the $0\%$ configuration appears as a maximum of that curve and above the rutile relaxed case.

The comparison of the two curves shows that the cell relaxation is not a by-product of the transition but an essential part of it. 
The strain couples to the square of the order parameter, so that a strain of one sign lowers the energy whatever the handedness. 
This linear-quadratic coupling renormalizes the quartic coefficient of the potential and, when it is strong enough, deepens the wells and pushes the equilibrium amplitude to much larger values.
Hence, the amplitude of the chiral distortion is essentially set by the cell rather than by the $Z_4$ mode alone. 
Together with the $7.3\%$ volume change, this large coupling indicates a strongly first-order transition, consistent with the marked hysteresis and the wide metastability range reported experimentally for the $\gamma$ phase~\cite{Denes1980transitions}.

The same curves give the barriers for converting one enantiomorph into the other. 
Going back through the rutile reference costs $89$~\si{\milli\electronvolt}/atom, while the path through the achiral $Cmcm$ phase identified in the previous section costs $54$~\si{\milli\electronvolt}/atom. NEB calculations confirm that the minimum energy path between the two enantiomers goes through the achiral $Cmcm$ configuration. 
Even this energy cost remains large: with twelve atoms in the cell it amounts to about $0.6$~\si{\electronvolt} per cell, well above the thermal energy available at the temperatures at which the $\gamma$ phase exists ($k_BT \simeq 37$~\si{\milli\electronvolt} at $428$~K). 
A thermally activated interconversion of the handedness is therefore not expected in SnF$_2$ (a full molecular dynamics would be required to check it properly), and the $Cmcm$ route would have to be driven externally, for instance by the circularly polarized field mechanism proposed for K$_3$NiO$_2$~\cite{Fava2025handedness}. 
With the rutile cell parameters, the energy barrier through the rutile phase decreases to $62$~\si{\milli\electronvolt}/atom. 

%%%%%%%%%%%%%%%%%%%%%%%%%%%%%%%%%%%%%%%%%%%%%%%%%%%%%%%
\section{\texorpdfstring{$\bm{\mathrm{TeO}}_{2}$}{TeO2}}

Tellurium dioxide (TeO$_2$) has been widely studied for its optical properties, in particular in nonlinear optics~\cite{pathomas1988, jhajose2012, philipkim2024, kondratyuk1987}. 
Like SnF$_2$, it belongs to the {\it AX}$_2$ family and its cation carries a $5s^2$ lone pair, Te$^{4+}$ and Sn$^{2+}$ having the same [Kr]\,$4d^{10}5s^2$ configuration. 
It also exists in three polymorphs: $\alpha$-TeO$_2$ (paratellurite, {\it P$4_12_12$} (92) or {\it P$4_32_12$} (96)), $\beta$-TeO$_2$ (tellurite, {\it Pbca} (61)) and $\gamma$-TeO$_2$ ({\it P$2_12_12_1$} (19))~\cite{Mirgorodsky2000, champarnaud2000}, all built from TeO$_4$ units with different connectivities~\cite{yfancheng2010}. 
Each of them has a counterpart in SnF$_2$, but their stability and the way they are obtained are different.

Paratellurite has the same enantiomorphic symmetry as $\gamma$-SnF$_2$, with $a = 4.8082$~\AA{} and $c = 7.6212$~\AA~\cite{pathomas1988}. 
Each Te atom is bonded to two equatorial oxygens at $1.879$~\AA{} and two axial oxygens at $2.121$~\AA{}, which forms a TeO$_4$ disphenoid, also described as a TeO$_4$E trigonal bipyramid where the lone pair E occupies the third equatorial site~\cite{yfancheng2010, Sabadel1997}. 
Contrary to $\gamma$-SnF$_2$, which is stable only between $428$ and $486$~K, $\alpha$-TeO$_2$ is the thermodynamically stable form~\cite{Deringer2014} and is grown as large single crystals for acousto-optic applications~\cite{kondratyuk1987}.

In nature, however, TeO$_2$ is mostly found as the mineral tellurite, $\beta$-TeO$_2$~\cite{Deringer2014}, whose orthorhombic structure is made of layers built from pairs of edge-sharing TeO$_4$ units~\cite{Beyer1967, Sabadel1997}, in contrast with the three-dimensional network of paratellurite. 
Since there is no group-subgroup relation between the centrosymmetric $Pbca$ group and $P4_12_12/P4_32_12$, this transformation is reconstructive, like the $\alpha \rightarrow \gamma$ transition of SnF$_2$, and it also does not provide a displacive path to chirality. 
The metastable $\gamma$-TeO$_2$ is obtained by devitrification of pure TeO$_2$ glass around 390$^\circ$C~\cite{Mirgorodsky2000, yfancheng2010}. 
It has the $P2_12_12_1$ symmetry of $\beta$-SnF$_2$, but a more distorted structural unit, with one Te--O bond much longer than the three others (TeO$_{3+1}$ unit), and it is formed from the glass rather than by cooling the chiral phase as in SnF$_2$.
Several other phases have been reported under pressure but they are not relevant for the present work~\cite{Peercy1974, Worlton1975, Skelton1976,Jayaraman1991, philipkim2024}.

The two compounds also differ by their rutile phase. 
It has never been observed for SnF$_2$, whereas a rutile TeO$_2$ film has been grown on a FeTe surface~\cite{Peng2022teo2}. 
As for SnF$_2$, we take the rutile $P4_2/mnm$ (136) structure as the high-symmetry reference from which we compute the phonons and the energies of the distorted phases of TeO$_2$.

\subsection{Phonon dispersion of Rutile high symmetry reference structure}
We relaxed TeO$_2$ within the same rutile $P4_2/mnm$ symmetry as SnF$_2$, which gives $a = 5.062$~\AA{} and $c = 3.398$~\AA, in agreement with the values obtained with the same functional in Ref.~\cite{GomezOrtiz2025pathways} ($a = 5.060$~\AA{} and $c = 3.401$~\AA).
The phonon dispersion curves are reported in Fig.~\ref{fig:TeO2_phonon} and the frequencies and irreps of the unstable modes in Tab.~\ref{tab:phonon_instable_points}.

The phonon dispersion curves of rutile TeO$_2$ are very similar to those of SnF$_2$ (Figs.~\ref{fig:snf2_phonon} vs \ref{fig:TeO2_phonon}), the same unstable modes, with the same irreps, are found at the same high-symmetry points (Tab.~\ref{tab:phonon_instable_points}).
The main difference is the amplitude of the instabilities, the imaginary frequencies being from about $1.5$ ($X_1$) to $3$ ($M_5^-$) times larger in TeO$_2$ than in SnF$_2$ (Tab.~\ref{tab:phonon_instable_points}).
In both compounds, the three strongest instabilities are found at the $Z$, $R$ and $A$ points and the lowest branch is almost flat along the $A-R-Z$ path.

Unlike what was found in SnF$_2$, the strongest instability of the TeO$_2$ whole spectrum is found at $Z_4$, the mode of interest for the present work. This mode is the doubly degenerate at the $Z = (0,0,\frac{1}{2})$ zone-boundary point and it is the only instability of the spectrum that can lead to the $P4_12_12/P4_32_12$ enantiomorphic pair, and its double degeneracy is what allows both handedness to be generated from a single unstable mode.  
This $Z_4$ mode is found at $363i$~\si{\per\centi\meter}, in good agreement with the $370i$~\si{\per\centi\meter} of Ref.~\cite{GomezOrtiz2025pathways}, and it is the strongest instability of the TeO$_2$ spectrum, slightly above the $R_1^+$ mode at $356i$~\si{\per\centi\meter} while the order of these two modes is opposite in SnF$_2$.

As for SnF$_2$, the imaginary frequencies do not give the order of the energies of the related phases, and each instability has to be condensed and relaxed, which is the object of the next section.

\subsection{Phonon condensation and phase diagram}
\begin{figure}[h!]
\centering
\begin{tikzpicture}[
    x=0.7cm, %pour ajuter l'etirement et la compression
    y=-0.05cm, 
    gpbox/.style={draw, fill=white, inner sep=2pt, font=\scriptsize, minimum height=1.2em},
    redbox/.style={gpbox, draw=red!70!black},
    greenbox/.style={gpbox, draw=green!50!black},
    lbl/.style={blue!70!black, font=\scriptsize\itshape, node distance=0pt},
]

    % AXE VERTICAL AVEC COMPRESSION
    \draw[thick, -{Stealth}] (0,-3) -- (0,200) node[below, align=left, font=\tiny] {Energy Gain\\(meV/atom)};
    
    % --- GRADUATIONS ET LIGNES (COORDONNÉES RECALCULÉES) ---
    % Le 0
    \draw[thick] (-0.1, 0) -- (0.1, 0);
    \node[left=2pt, font=\tiny\bfseries] at (0, 0) {0};
    \draw[thick] (0,0) -- (10,0); % Ligne du haut (Rutile)

    % Fonction de mapping pour remonter le bloc : 
    \foreach \val/\yplot in {20, 40, 60, 80, 100, 120, 140, 160, 180, 200} {
        \draw[thick] (-0.1, \yplot) -- (0.1, \yplot);
        \node[left=2pt, font=\tiny\bfseries] at (0, \yplot) {\val};
        \draw[dotted, black!40] (0.1, \yplot) -- (10, \yplot); 
    }

    % ELÉMENTS DU DIAGRAMME
    
    % Niveau 0
    \node[gpbox] at (5,0) {$P4_2/mnm$ (136)};

    % BLOC DU BAS 
    \node[gpbox] (p113) at (1.5,150) {$P\bar{4}2_1m$ (113)};
    \node[lbl, above=of p113] {$\Gamma_2^-(a)$};
    
    \node[gpbox] (p31) at (1.5,93) {$Pmn2_1$ (31)};
    \node[lbl, above=of p31] {$\Gamma_5^-(a,0)$};

    \node[gpbox] (p38) at (1.5,39) {$Amm2$ (38)};
    \node[lbl, above=of p38] {$\Gamma_5^-(a,a)$};
    
    \node[gpbox] (p13) at (9,8) {$P2/c$ (13)};
    \node[lbl, left=of p13] {$X_1(a,a;0,0)$};

    \node[gpbox] (p33) at (1.5,65) {$Pna2_1$ (33)};
    \node[lbl, above=of p33] {$X_1 (a,0;0,0)$};

    \node[gpbox] (c63) at (9,67) {$Cmcm$ (63)};
    \node[lbl, below=of c63] {$Z_4(a,0)$};
    
    \node[redbox] (p92) at (5,180) {$P4_{1/3}2_12$ (92/96)};
    \node[lbl, above=of p92] {$\alpha, Z_4(a, \pm a)$};

     \node[gpbox] (p59) at (5,131) {$Pmnm$ (59)};
    \node[lbl, above=of p59] {$M_2^- M_3^-(a,b)$};

    \node[gpbox] (p84) at (5,69) {$P4_2/m$ (84)};
    \node[lbl, above=of p84] {$M_5^-(a,0)$};
    
    \node[gpbox] (p51) at (9,33) {$Pmm2$ (51)};
    \node[lbl, above=of p51] {$M_5^-(a,a)$};

    \node[gpbox] (p12) at (9,56) {$C2/m$ (12)};
    \node[lbl, above=of p12] {$R_1^+(a,a;0,0)$};

    \node[gpbox] (p14) at (9,179) {$P2_1/c$ (14)};
    \node[lbl, above=of p14] {$R_1^+(a,0;0,0)$};
    
    \node[gpbox] (i74) at (5,46) {$Imma$ (74)};
    \node[lbl, above=of i74] {$A_4(a,0)$};

    \node[gpbox] (i122) at (9,128) {$I\bar{4}2d$ (122)};
    \node[lbl, above=of i122] {$A_4(a,a)$};

    \node[greenbox] (p19) at (1.5,167) {$P2_12_12_1$ (19)};
    \node[lbl, above=of p19] {$\gamma$};

    \node[greenbox] (p61) at (1.5,184) {$Pbca$ (61)};
    \node[lbl, above=of p61] {$\beta$};

\end{tikzpicture}
\caption{Same as Fig.~\ref{fig:snf2_phase_diagram} for TeO$_2$: hierarchy of the energy gains with respect to the rutile parent phase, in \si{\milli\electronvolt}/atom.
For each phase, the space group and the main condensing phonon mode (in blue) are given. 
The red box highlights the chiral phases which have the largest energy gain among the phases derived from the rutile instabilities. 
The green boxes highlight the $\gamma-$ and $\beta-$ TeO$_2$ phases, observed experimentally. They respectively have an energy gain of -167 and -184 \si{\milli\electronvolt}/atom. 
}
\label{fig:teo2_phase_diagram}
\end{figure}
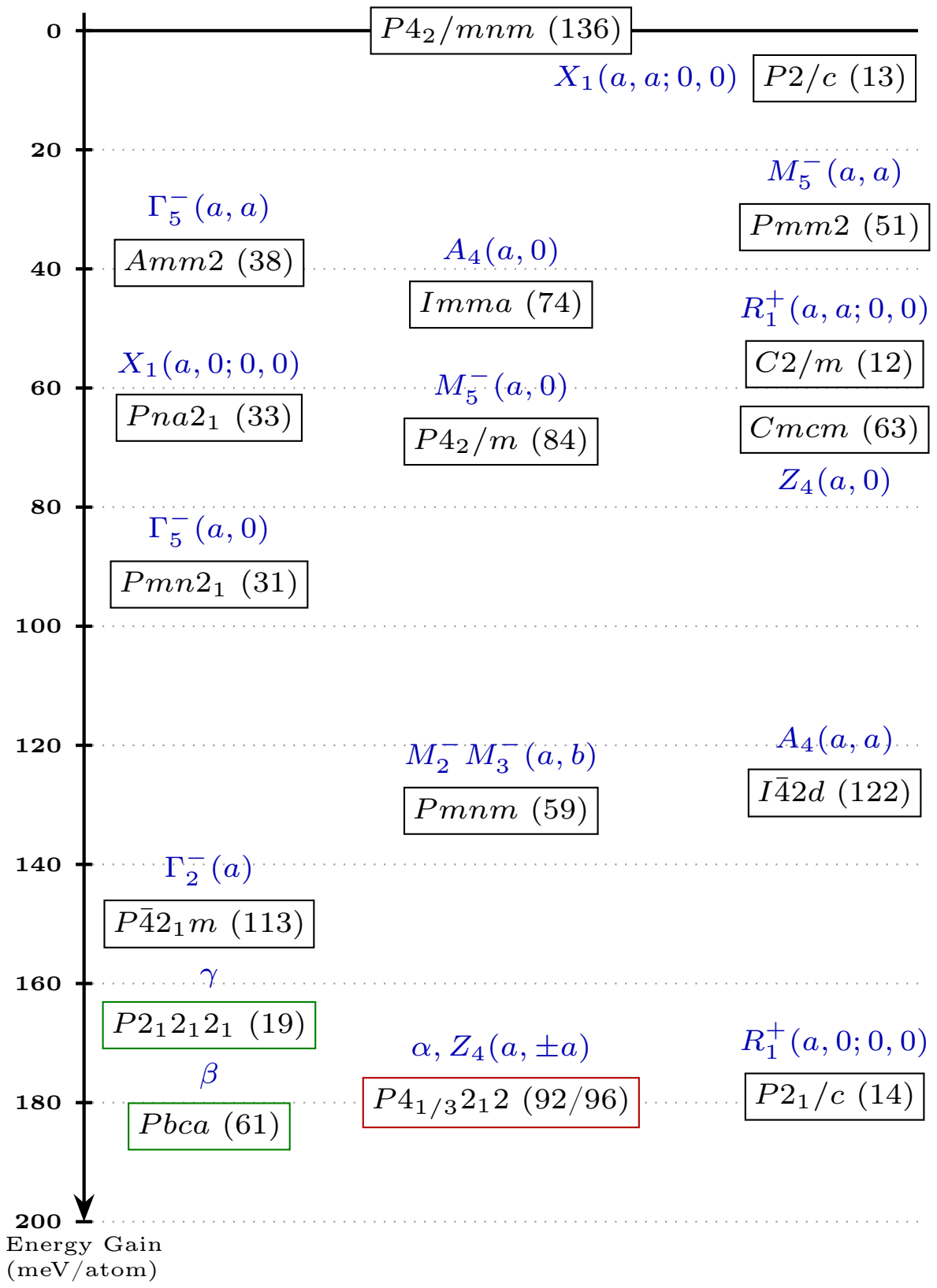

The unstable modes of TeO$_2$ were condensed and relaxed following the same procedure as for SnF$_2$.
The resulting energies are given in Tab.~\ref{tab:phonon_instable_points} and their order is summarized in Fig.~\ref{fig:teo2_phase_diagram}, to be compared with Fig.~\ref{fig:snf2_phase_diagram}.

As for SnF$_2$, the main observation is that the chiral phases are the most stable of the phases derived from the rutile instabilities.
Condensing the doubly degenerate $Z_4$ mode along the diagonal $(a,a)$ and $(a,-a)$ directions gives the $P4_12_12$ (92) and $P4_32_12$ (96) phases with an energy gain of $180$~\si{\milli\electronvolt}/atom, the largest of all the instabilities considered, which is about twice the gain obtained in SnF$_2$ ($89$~\si{\milli\electronvolt}/atom).
The chiral phases of TeO$_2$ can therefore also be connected to the rutile reference through a single unstable phonon mode, and they also correspond to the deepest minimum that can be reached from it.

The non-diagonal directions of the same $Z_4$ mode lead to a different phase. 
When only one component of the order parameter is condensed, i.e. along $(0,a)$, the system relaxes into the achiral orthorhombic $Cmcm$ (63) phase with an energy gain of $67$~\si{\milli\electronvolt}/atom. 
As described in SnF$_2$, the role of this phase is not that of a competing ground state but that of a transition state, i.e. the achiral configuration through which the two handedness can be interconverted without going back to the much higher rutile reference.
The associated barrier, $113$~\si{\milli\electronvolt}/atom (twice the $54$~\si{\milli\electronvolt}/atom of SnF$_2$), is discussed below together with the energy wells.
The ratio between the energy gains of the $Cmcm$ and chiral phases is almost the same in the two compounds ($0.39$ in SnF$_2$ and $0.37$ in TeO$_2$), which indicates that the energy surface of the $Z_4$ order parameter has the same shape in both of them, with an energy scale about twice larger in TeO$_2$.

The chiral phases are, however, not isolated in energy.  
As in SnF$_2$, another phase lies within $1$~\si{\milli\electronvolt}/atom of them, but it does not come from the same instability.
The $R_1^+$ instability relaxes into a monoclinic $P2_1/c$ (14) phase lying only $1$~\si{\milli\electronvolt}/atom above the chiral phases (against $5$~\si{\milli\electronvolt}/atom in SnF$_2$), while the $Pna2_1$ phase coming from the $X_1$ mode lies here $115$~\si{\milli\electronvolt}/atom higher.
These two phases are essentially degenerate within the accuracy that can be expected from the exchange-correlation functional, so that the present calculations cannot tell which of the two is the lowest-energy phase derived from the rutile reference. 
More generally, the energy gains are $1.4$ to $2.5$ times larger in TeO$_2$ than in SnF$_2$ for all the phases except the $Pna2_1$ ($X_1$), $Pmm2$ ($M_5^-$) and $C2/m$ ($R_1^+$) ones, whose gains are smaller than or close to those of SnF$_2$.
This is why the closest competitor of the chiral phases changes from $Pna2_1$ in SnF$_2$ to $P2_1/c$ in TeO$_2$.
The remaining instabilities are clearly higher in energy, between $8$ and $150$~\si{\milli\electronvolt}/atom below the parent phase, i.e. at least $30$~\si{\milli\electronvolt}/atom above the chiral phases. 
In particular, the polar $Pmn2_1$ phase coming from the $\Gamma_5^-$ mode lies $87$~\si{\milli\electronvolt}/atom above the chiral phases (against $37$~\si{\milli\electronvolt}/atom in SnF$_2$) such that the rutile phase of TeO$_2$ has a marked preference for the chiral distortion over the ferroelectric one.

As for SnF$_2$, we have also relaxed the experimentally reported phases (green boxes in Fig.~\ref{fig:teo2_phase_diagram}).
The metastable $\gamma$-TeO$_2$ ($P2_12_12_1$) lies $13$~\si{\milli\electronvolt}/atom above the chiral $\alpha$ phase.
The tellurite $\beta$-TeO$_2$ ($Pbca$) is found $4$~\si{\milli\electronvolt}/atom below the chiral phases.
This is the same situation as in SnF$_2$, where the achiral $\alpha$ phase ($C2/c$) lies $5$~\si{\milli\electronvolt}/atom below the chiral one but in TeO$_2$ it does not agree with experiment: calorimetry shows that $\alpha$-TeO$_2$ is the most stable polymorph, $\beta$-TeO$_2$ being higher in enthalpy by $1.4$~kJ/mol, i.e. about $5$~\si{\milli\electronvolt}/atom~\cite{Majzlan2022}.
The same inversion is obtained with the PBE functional~\cite{Ceriotti2006teo2}, and the experimental ordering is recovered when suitable dispersion corrections are included~\cite{Deringer2014}.

The SAM decomposition gives the same secondary modes as in SnF$_2$ for all the phases (last column of Tab.~\ref{tab:phonon_instable_points}).
For the chiral phases, $\Gamma_1^+$ is again the only secondary mode, i.e. the distortion from the rutile phase to $P4_12_12/P4_32_12$ is entirely described by the $Z_4$ order parameter together with the symmetry-preserving
relaxation of the cell, with no additional symmetry-adapted distortion, so that the two contributions can be separated as for SnF$_2$

%%%%%%%%%%%%%%%%%%%%%%%%%%%%%%%%%%
\subsection{Energy wells of the chiral phonon mode}

The SAM analysis has shown that the distortion connecting the rutile reference to the chiral phases involves the same two main ingredients as in SnF\(_2\): the $Z_4$ order parameter and the fully symmetric relaxation of the cell.  
We have therefore computed the energy as a function of the $Z_4$ amplitude for the same two choices of cell.

In the first case, the lattice parameters are constrained to those of the optimized rutile parent phase doubled along the $z$ axis, $a = b = 5.062$~\AA{} and $c = 6.796$~\AA, and only the atoms are displaced along the $Z_4$ eigenvector. 
In the second case, the lattice parameters are those of the fully relaxed chiral structure, $a = b = 4.839$~\AA{} and $c = 7.434$~\AA . These values agree with the experimentally measured lattice parameters, $a = 4.8082$~\AA{} and $c = 7.6212$~\AA~\cite{pathomas1988}, $a$ being overestimated by $0.6\%$ and $c$ underestimated by $2.5\%$.
As in SnF$_2$ ($3.6\%$), $c$ is the parameter that is underestimated. 
In both cases, the amplitude is normalized so that $\pm 100\%$ corresponds to the atomic positions of the two enantiomers for the cell under consideration, and $0\%$ to the rutile atomic positions.
The two cells differ considerably: going from the rutile to the chiral one corresponds to a contraction of $4.4\%$ in the $ab$ plane and an elongation of $9.4\%$ along $c$, i.e. a change of the $c/a$ ratio from $1.343$ to $1.536$ and a reduction of the volume by only $0.04\%$. 
Contrary to SnF$_2$, where the volume is reduced by $7.3\%$, the volume is thus conserved, the in-plane contraction being compensated by the elongation along $c$.
The change of the $c/a$ ratio is, however, similar in the two compounds ($+14\%$ in TeO$_2$ and $+16\%$ in SnF$_2$).
The strain that accompanies the $Z_4$ distortion in both compounds is therefore mainly the tetragonal elongation of the cell, the volume contraction being specific to SnF$_2$.
    
The energies obtained in those two cases are reported in Fig.~\ref{fig:teo2_energydiff_phase}.
Both curves are similar to those obtained for SnF\(_2\), showing the double-well shape expected from an unstable phonon mode, with the two minima strictly symmetric with respect to the achiral $0\%$ configuration, the energy scale being about twice larger in TeO$_2$.
At the rutile cell (red curve), the $Z_4$ distortion is unstable and the minimum ($-125$~\si{\milli\electronvolt}/atom) is reached for a displacement of the tellurium atoms of $\pm 0.104$~\AA. 
At the chiral cell (blue curve), the double well is deeper ($-180$~\si{\milli\electronvolt}/atom) and it is reached for slightly smaller tellurium displacements of $\pm 0.083$~\AA.
In addition, the blue curve does not start from zero, imposing the chiral cell with the atoms kept at their rutile positions costs about $69$~\si{\milli\electronvolt}/atom, against about $53$~\si{\milli\electronvolt}/atom in SnF$_2$.
This elastic cost is larger in TeO$_2$ even though its volume is conserved, which reflects the stiffer lattice of the oxide.

The comparison with SnF$_2$ shows that the cell relaxation does not play exactly the same role in the two compounds.
In both of them, the energy gained at the rutile cell is about $70\%$ of the total one ($125$ out of $180$~\si{\milli\electronvolt}/atom in TeO$_2$ and $62$ out of $89$~\si{\milli\electronvolt}/atom in SnF$_2$), the remaining $30\%$ coming from the cell relaxation through the coupling between the strain and the square of the order parameter.
In SnF$_2$, however, the tin displacement increases by a factor of seven from the rutile to the chiral cell, so that the amplitude of the chiral distortion is set by the cell.
In TeO$_2$, the tellurium displacement is instead slightly reduced (from $0.103$ to $0.083$~\AA): the $Z_4$ mode is unstable enough to reach a large amplitude by itself at the rutile cell, and the cell relaxation mainly deepens the wells without increasing the cation displacement.
In addition, as the volume is conserved, the coupling involves the tetragonal strain only.
The volume change that was taken as an indication of a strongly first-order transition in SnF$_2$ is thus absent in TeO$_2$, although the large change of $c/a$ still points toward a first-order character.

The same curves give the barriers for converting one enantiomorph into the other. 
Returning through the rutile reference costs $180$~\si{\milli\electronvolt}/atom, while the path through the achiral $Cmcm$ phase identified in the previous section costs $113$~\si{\milli\electronvolt}/atom, i.e. about twice the SnF$_2$ values ($89$ and $54$~\si{\milli\electronvolt}/atom). With twelve atoms in the cell, the barrier through the $Cmcm$ phase amounts to about $1.4$~\si{\electronvolt} per cell, which is much larger than the thermal energy even at the melting temperature of paratellurite ($k_BT \simeq 86$~\si{\milli\electronvolt} at about $1000$~K).
A thermally activated interconversion of the handedness is thus even less likely in TeO$_2$ than in SnF$_2$.

If the rutile cell parameters are imposed while allowing the atomic positions to relax, the barrier through the rutile reference is reduced to $125$~\si{\milli\electronvolt}/atom. 
Despite the energy gain associated with the chiral distortion, the rutile phase can nevertheless be stabilized in thin-film form. 
Epitaxial growth on a tetragonal FeTe surface has indeed been reported to stabilize a rutile-structured TeO$_2$ film through interfacial interactions. 
This shows that substrate-induced strain and interface engineering can suppress the bulk instability.~\cite{Peng2022teo2}

%%%%%%%%%%%%%%%%%%%%%%%%%%%%%%%%%%%%%%%%%%%%%%%%%%%
\section{Nature of the driving forces}

The rutile phase of SnF$_2$ and TeO$_2$ carries a chiral instability at $Z$, whereas rutile SnO$_2$ or PbO$_2$ do not. 
To identify what controls the presence of this $Z_4$ mode, we have computed the phonon frequencies of a series of AX$_2$ compounds in the same rutile structure. 
With the exception of SnO$_2$, GeO$_2$ and PbO$_2$, this structure is hypothetical for most of them and is used here only as a common reference allowing the different chemistries to be compared at fixed symmetry.

\subsection{Lone pair presence as a condition}
The first hypothesis is that the lone pair of electrons possessed by the cation causes the instability. Indeed, lone pairs are known to have an anisotropic electronic density and to break the symmetry of the local environment \cite{Walsh2011lonepair}.\\
Our comparative study was done on different crystal belonging in the rutile \(AX_2\), summarized in Tab. \ref{tab:lone_pair_presence}. They are separated into two categories, whether the cation possess a lone pair of electrons or not.

\begin{table}[htbp]
\centering
\renewcommand{\arraystretch}{1.3}
\begin{tabular}{l c l c}
\textbf{Compound} & \textbf{Cations} &\textbf{Valence Configuration}  & \textbf{Lone Pair?} \\
\hline
 SnF\(_2\), SnCl\(_2\) & Sn\(^{2+}\) & [Kr] 4d$^{10}$ \textbf{5s$^2$} 5p$^0$ & Yes \\
 SnO\(_2\) & Sn\(^{4+}\) & [Kr] 4d$^{10}$ 5s$^0$ 5p$^0$ & No \\
 GeF\(_2\),GeCl\(_2\) & Ge\(^{2+}\) & [Ar] 3d$^{10}$ \textbf{4s$^2$} 4p$^0$ & Yes \\
 GeO\(_2\) & Ge\(^{4+}\) & [Ar] 3d$^{10}$ 4s$^0$ 4p$^0$ & No \\
 PbF\(_2\), PbCl\(_2\) & Pb\(^{2+}\) & [Xe] 4f$^{14}$ 5d$^{10}$ \textbf{6s$^2$} 6p$^0$ & Yes \\
 PbO\(_2\) & Pb\(^{4+}\) & [Xe] 4f$^{14}$ 5d$^{10}$ 6s$^0$ 6p$^0$ & No \\
 SeO\(_2\) & Se\(^{4+}\) & [Ar] 3d$^{10}$ \textbf{4s$^2$} 4p$^0$ & Yes \\
 TeO\(_2\) & Te\(^{4+}\) & [Kr] 4d$^{10}$ \textbf{5s$^2$} 5p$^0$ & Yes \\\hline
\end{tabular}
\caption{Valence configurations of the cations in selected {\it AX$_2$} compounds and the verification of the presence of a stereochemically active lone pair.}
\label{tab:lone_pair_presence}
\end{table}

For each compound, phonon frequencies were calculated, and more specifically at $Z \; (0 \, 0 \,\frac{1}{2} )$, which is the q-point of interest. The results are summarized in Tab. \ref{tab:lone_pair_frequency}. 
We observe first that none of the compounds whose cation has no lone pair develops an instability of the $Z_4$ mode. In contrast, almost all lone-pair compounds exhibit the instability (imaginary frequency), with the single exception of PbCl$_2$. 
This correlation could at first be attributed to the anion or to the formal charge of the cation rather than to the lone pair itself, since the lone-pair compounds of the series are mostly halides with a $2+$ cation. 
The oxides alone, however, already discriminate between the two: SeO$_2$ and TeO$_2$, with a lone pair, are unstable, while SnO$_2$, GeO$_2$ and PbO$_2$, with the same anion and the same $4+$ formal charge but no lone pair, are not. 
The presence of the lone pair is therefore the relevant parameter.
The case of PbCl$_2$ shows at the same time that it is not a sufficient condition. We analyze in the following what else is required.

\begin{table}[htbp]
    \centering
    \renewcommand{\arraystretch}{1.5}
    \begin{tabular}{c |c|c|c}
       \textbf{ Compound }& \textbf{Lone}& \textbf{Z\(_4\) frequency} & \textbf{\(\Delta E_{\text{parent-daughter}} \)} \\
        \textbf{ AX\(_2\)}& \textbf{ Pair ?} & (\si{\per\centi\meter}) &(meV/atom)\\\hline
         SnF\(_2\)& Yes & $154i$ & -89\\
         SnCl\(_2\)& Yes & $43i$ &-1\\
         SnO\(_2\)& No & $125$ &- \\
         GeF\(_2\)& Yes & $197i$ & -140\\
         GeCl\(_2\)& Yes & $62i$& -13\\
         GeO\(_2\)& No & $190$ &-\\
         PbCl\(_2\)& \textbf{Yes} & $\mathbf{20}$ & -\\
         PbF\(_2\)&Yes & $51i$ &-15\\
         PbO\(_2\)& No & $59$ & - \\
         SeO\(_2\)& Yes & $382i$& -211\\
         TeO\(_2\)& Yes & $363i$& -180
    \end{tabular}
    \caption{Correlation between lone pair presence and $Z_4$ phonon instability in {\it AX$_2$}. An imaginary frequency correspond to an instability of the mode. \(\Delta E_{\text{parent-daughter}} \) is the stabilization of the daughter phase taking the rutile phase as a reference.
    }
    \label{tab:lone_pair_frequency}
\end{table}

\subsection{Pseudo Jahn-Teller Effect}

To understand the different behaviours in {\it AX$_2$} compounds, it is necessary to go further than the simple lone pair model \cite{Walsh2011lonepair}. The refined model based on the orbital interaction, and not simply the steric volume, identifies the driving force as a pseudo Jahn-Taller effect (PJTE), also called second order Jahn-Teller effect \cite{Bersuker2021pjte}.\\

In the rutile structure the cation occupies the $2a$ Wyckoff position, whose site symmetry $mmm$ is centrosymmetric. On-site mixing between the cation $ns$ and $np$ orbitals, which have opposite parity, is therefore forbidden, and the lone pair remains stereochemically inactive. Any mechanism making the lone pair active thus requires the removal of the inversion centre at the cation site.

The chiral phonon mode $Z_4$ corresponds to a distortion that breaks the local inversion symmetry. This specific structural displacement is the vibronic coupling needed to lift the forbidden mixing of $ns-np$ orbital. To understand the electronic origin of the transition, we model the mechanism through interactions. Initially, the interaction between cation $s$ orbital and anion $p$ orbital states create occupied anti-bonding states \((s-p)^*\) at the top of the valence band. The structural distortion $Z_4$ then breaks the local inversion symmetry at the cation site. It allows the hybridization between \((s-p)^*\) states and the nominally empty cation $p$ orbitals in the conduction band. Finally, this coupling results in a lowering of the total electronic energy and a stabilization of a lower-symmetry chiral structure.

\begin{figure}
    \centering
    \includegraphics[width=0.48\linewidth]{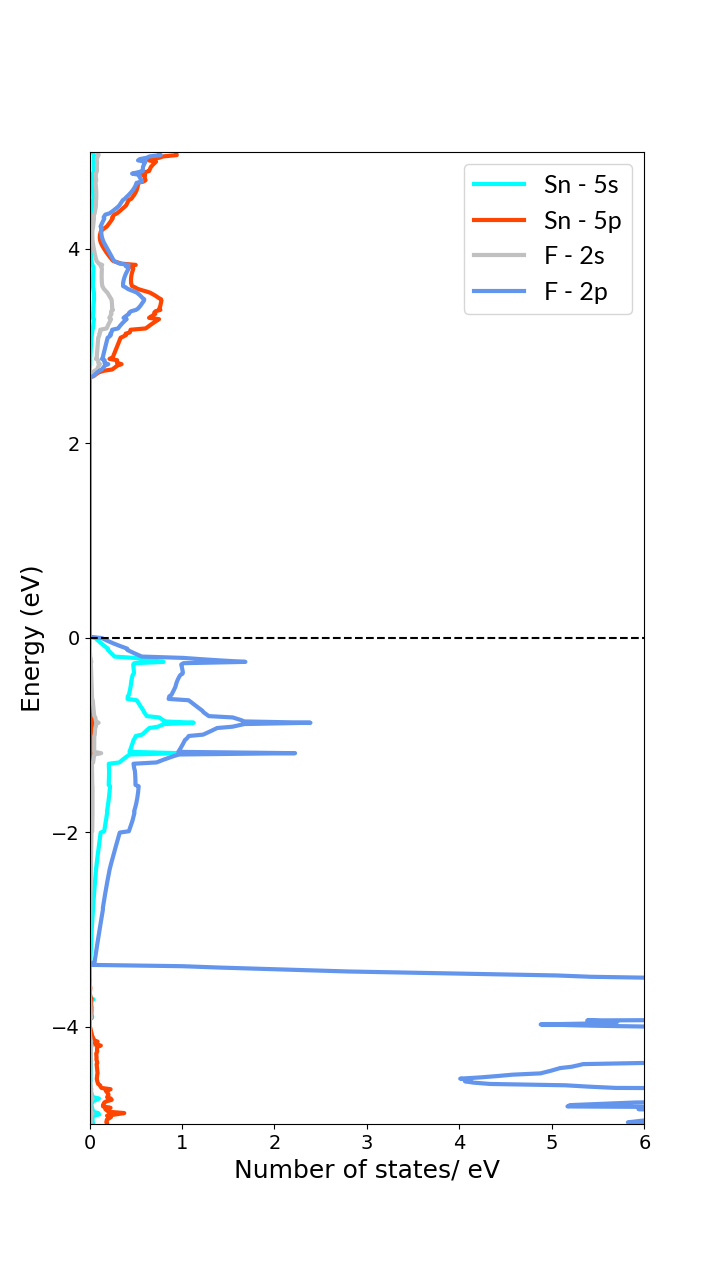}
    \includegraphics[width=0.48\linewidth]{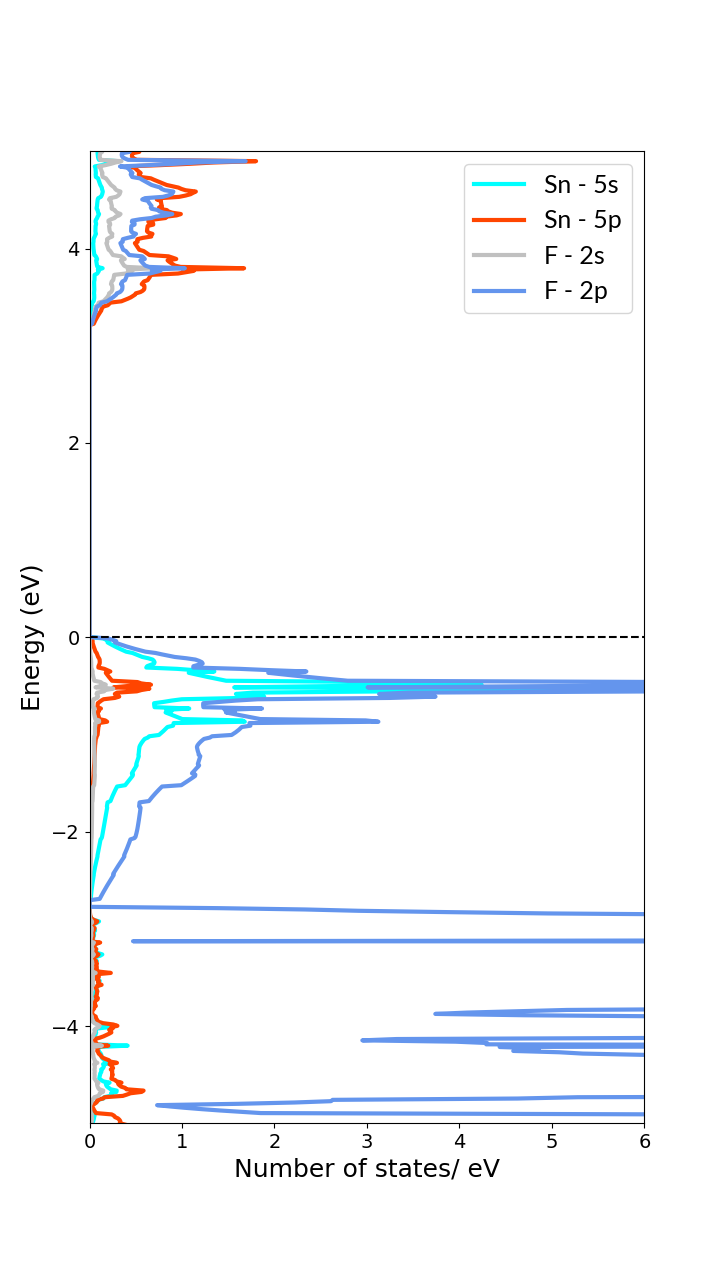}
    \caption{Density of states associated with SnF\(_2\) a) in rutile $P4_2/mnm$ and b) chiral $P4_{1/3}2_12$ with a focus close to the electronic bandgap. In the first case, the top of the valence band is occupied with the anti-bonding states formed by Sn\(-5s\) and F\(-2p\) states, respectively in cyan and blue. There is no presence of Sn$-5p$ states, indicated in red, forbidden by inversion symmetry. The bandgap has a value of $2.51$ eV. After the $Z_4$ distortion, the top of the valence band manifests the presence of those Sn$-5p$ states. This is the sign of a hybridization. The bandgap increases to $3.11$ eV, showing that the structure is electronically more stable. }
    \label{fig:dos_snf2}
\end{figure}

This mechanism is illustrated with in Fig. \ref{fig:dos_snf2} for tin fluoride. It represents the Density of States (DOS) associated with the electronic band structure of SnF$_2$ of the parent rutile and chiral daughters phases. In the rutile phase (see Fig. \ref{fig:dos_snf2}a), the valence band is dominated by Sn$-s$ and F$-2p$ anti-bonding states. The Sn$-5p$ states (in red) are absent due to inversion symmetry. The signature of the mechanism is the appearance of Sn$-5p$ weight at the top of the valence band in the chiral phase, which is forbidden by symmetry in the rutile one and is therefore a direct symmetry-allowed fingerprint of the $s$-$p$ mixing. 
It is accompanied by an opening of the electronic band gap from $2.51$ to $3.11$~\si{\electronvolt}, consistent with a stabilization of the occupied states. We note that both values are obtained at the PBEsol level and are underestimated with respect to experiment, so that only their relative change is meaningful here.

\subsection{Competition for the instability}

The absence of instability in PbCl\(_2\) demonstrates that the presence of a lone pair of electrons is a necessary but insufficient condition for the transition. The outcome depends on a competition between electronic stabilization and lattice resistance.\\
Following Bersuker's framework \cite{Bersuker2013formula,Bersuker2021pjte}, the energy curvature $K$, proportional to the square of the phonon frequency $\omega^2$, can be expressed as :
\begin{align}
    \label{eq:bersuker_lonepair}
    K&=K_0-\frac{2F^2}{\Delta}
\end{align}
, where $K_0$ represents the lattice rigidity (resistance to external deformation), $F$ is the vibronic coupling force (PTJE force) and $\Delta$ the energy separation between the two vibronically coupled states. In this system, $\Delta$ corresponds to the energy gap between the occupied anti-bonding  \((s-p)^*\) states at the top of the valence band and the nominally empty cation $np$ states in the conduction band.
For an achiral-to chiral phase transition to occur in this case, the energy curvature must be negative, $K<0$. This leads to the following condition:

\begin{align}
    \label{eq:pjte_competition}
    K_0<\frac{2F^2}{\Delta}.
\end{align}

\begin{figure}[htbp]
    \centering
    \includegraphics[width=0.94\linewidth]{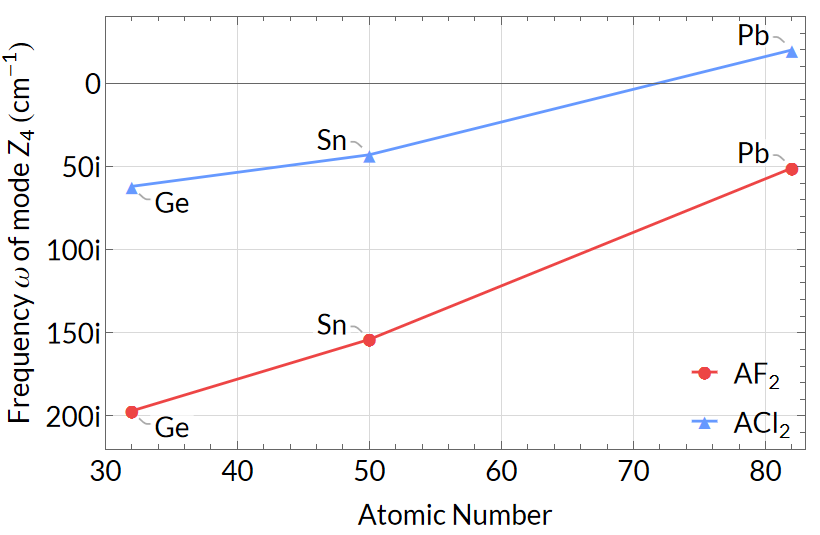}
    \caption{Phonon frequency associated with for chiral phonon mode $Z_4$ for six AX \(_2\) compounds. The $x$-axis corresponds to the atomic number associated with the cation in each compound. The red line indicates that the anion is fluorine and the blue, chlorine. 
    }
    \label{fig:z4_atomic}
\end{figure}

The structural transition occurs when the electronic destabilization overcome the lattice rigidity. Whether the inequality (Eq. \eqref{eq:pjte_competition}) is satisfied depends on the chemical nature of the compounds. To investigate the sensitivity of this competition to ionic species, we plotted the phonon frequency of $Z_4$ mode for six studied compounds, as displayed in Fig.\ref{fig:z4_atomic}. For a given anion (on a same line), the instability decreases as the cation size increases. This is explained by an increase in the lattice rigidity $K_0$: a larger cation takes more place in the octahedra formed by the anion, leading to less place for atoms to move. Similarly, for a given cation, replacing fluorine with chlorine reduces the instability. This is explained by the same reasons as for the cation. Furthermore, the electronegativity that decreases, lowering the vibronic coupling $F$.

\begin{figure}[htbp]
    \centering
    \includegraphics[width=0.94\linewidth]{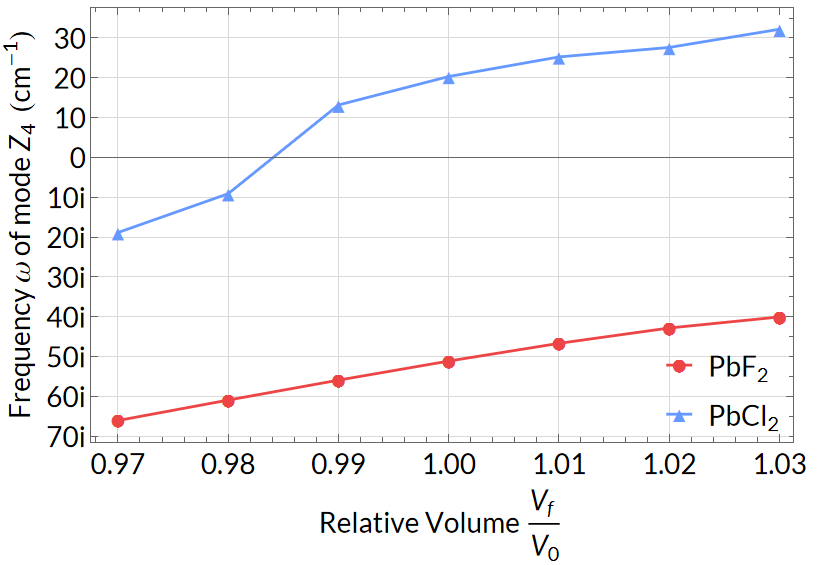}
    \caption{Evolution of the chiral phonon frequency $Z_4$ associated with homogeneous change of the volume cell in PbCl\(_2\) and PbF\(_2\). The phonon frequency of PbCl\(_2\) becomes imaginary with homogenous compression of about $1.6$\%. Imaginary frequencies are plotted as negative values for clarity. }
    \label{fig:pb_z4_volume}
\end{figure}

In PbCl\(_2\), the combination of a small coupling $F$ and a large bandgap $\Delta$ is not able to overcome the large rigidity $K_0$, making it stable. This explanation is demonstrated by studying the effect of an homogeneous volume change for both PbCl\(_2\) and PbF\(_2\), illustrated in Figure \ref{fig:pb_z4_volume}.

The two pictures make opposite predictions under hydrostatic strain. 
If the instability were of steric origin, expanding the cell would free the volume required by the lone pair and would favour the distortion. 
If it is driven by the vibronic coupling, compression should instead favour it, since reducing the interatomic distances increases the cation-anion orbital overlap and hence the coupling force $F$. 
As shown in Fig. \ref{fig:pb_z4_volume}, both compounds follow the same trend, corresponding to the second behaviour: the $Z_4$ instability softens under compression, while expansion hardens it. In PbF\(_2\), which is already unstable at equilibrium, the instability increases under pressure. In PbCl\(_2\), the $Z_4$ frequency becomes imaginary at about $1.6\%$. 
This occurs despite the fact that compression also stiffens the lattice and raises $K_0$. The fact that the electronic term $F/\Delta^2$ grows faster then the increasing lattice resistance $K$ shows that the chiral instability of these rutile compounds is therefore electronic in origin. It is controlled by orbital overlap, the volume acting on it only indirectly.

\section{Conclusion}
We have shown, through first-principles DFT calculations, that the chiral phases of SnF\(_2\) and TeO\(_2\) can be understood as arising from a displacive phase transition from a hypothetical rutile $P4_2/mnm$ parent phase.
In both compounds, we computed the phonon dispersion curves of this parent phase, performed a symmetry analysis of each unstable mode and condensed and relaxed each of them to build the hierarchy of the low-symmetry phases. This allowed us to identify the doubly degenerate $Z_4$ mode at the $Z$ zone-boundary point as the only instability able to generate the experimentally observed $P4_12_12/P4_32_12$ chiral phases. Condensation of this mode along the two diagonal directions $(a,a)$ and $(a,-a)$ produces the right- and left-handed structures, respectively.

The resulting energy landscape along the $Z_4$ path is a symmetric double well, with the achiral configuration in the centre and the two enantiomorphic configurations as equivalent minima. 
For SnF\(_2\), the chiral phases are stabilized by $89$~\si{\milli\electronvolt}/atom with respect to the rutile reference, whereas the corresponding stabilization reaches $180$~\si{\milli\electronvolt}/atom in TeO\(_2\). 
In both compounds, this is the largest energy gain among all the phases obtained from the rutile instabilities, with a single close competitor within $1$~\si{\milli\electronvolt}/atom ($Pna2_1$ in SnF$_2$ and $P2_1/c$ in TeO$_2$), while the polar $Pmn2_1$ phase lies $37$ and $87$~\si{\milli\electronvolt}/atom above the chiral phases, respectively. Despite the difference in the magnitude of the instability and of the energy gain, both compounds follow the same route toward chirality: the energy surface of the $Z_4$ order parameter has the same shape in the two compounds with an energy scale about twice as large in TeO$_2$.

Our analysis shows that the chiral distortion cannot be driven by the $Z_4$ atomic displacement alone. The simultaneous cell relaxation is the other essential ingredient: the lattice contracts in the $ab$ plane and expands along the $c$ axis.
This strain couples to the square of the order parameter, renormalizes the effective quartic term of the Landau potential and deepens the double well.
In SnF$_2$, the relaxation is associated with a large volume contraction (7.3\%), whereas in TeO$_2$ it occurs at nearly constant volume through an anisotropic redistribution of strain. 
However, in both cases, the structural distortion is large and the transition points toward a first-order transition rather than a small-amplitude continuous instability, the transition being expected to be strongly first-order in SnF$_2$, where it is accompanied by a large volume change. 

The same energy landscape also gives the energy barriers for switching the handedness. 
The direct path between the two enantiomorphs through the rutile parent phase requires $89$ and $180$~\si{\milli\electronvolt}/atom in SnF$_2$ and TeO$_2$, respectively. A lower-energy route is provided by the achiral $Cmcm$ phase obtained by condensing only one component of the $Z_4$ order parameter. 
This phase acts as a transition state rather than as a competing ground state, as confirmed by nudged elastic band calculations in SnF$_2$, reducing the switching barrier to $54$ and $113$~\si{\milli\electronvolt}/atom in SnF$_2$ and TeO$_2$, respectively. 
These barriers, about $0.6$ and $1.4$~\si{\electronvolt} per 12-atom cell, remain well above the thermal energy available near the stability range of the chiral phases, indicating that spontaneous thermal interconversion of handedness is unlikely. 

Beyond SnF$_2$ and TeO$_2$, we conducted a comparative study of rutile $AX_2$ compounds. This allowed us to establish a chemical criterion for the occurrence of this chiral instability : a stereochemically active $ns^2$ lone pair on the cation is necessary but not sufficient. 
All the compounds considered with a lone-pair free cation remain stable with respect to the $Z_4$ distortion, while all the lone-pair compounds except PbCl$_2$ develop the instability. 
The exception of PbCl$_2$ demonstrates that the electronic stabilization driven by the lone pair must also overcome the elastic rigidity of the lattice. 

The microscopic origin of this competition is a pseudo Jahn-Teller effect. 
In the centrosymmetric rutile structure, the local inversion symmetry at the cation site forbids on-site $ns$-$np$ hybridization. 
The $Z_4$ distortion breaks this local symmetry and allows the mixing between occupied anti-bonding states $(s-p)^*$ and the nominally empty cation $np$ states in the conduction band. 
The resulting electronic energy lowering competes with the lattice rigidity $K_0$, and the instability occurs when the vibronic stabilization exceeds this resistance. 
The appearance of cation-$p$ states at the top of the valence band in the chiral phase of SnF$_2$, together with the opening of the electronic gap,from $2.51$ to $3.11$~\si{\electronvolt}, provides a direct signature of this mechanism. 

The comparison of the frequency of the chiral $Z_4$ mode in the six halide compounds helps identify the parameters controlling the balance between electronic stabilization and lattice stiffness. 
For a given cation, replacing fluorine by chlorine weakens the instability, while increasing the cation size also decreases it. 
These observations are consistent with a reduction of the vibronic coupling and an increase of the elastic resistance to the distortion. 

Most importantly, the response of PbCl$_2$ and PbF$_2$ to hydrostatic compression shows that the $Z_4$ mode softens under pressure, even though compression also stiffens the lattice. 
This behaviour excludes a purely steric interpretation of the lone-pair effect and demonstrates that the instability is primarily controlled by cation-anion orbital overlap. 
In PbCl$_2$, a homogeneous compression of about $1.6\%$ is sufficient to turn the $Z_4$ frequency imaginary. This suggests that epitaxial strain or pressure could be used to tune compounds toward the chiral instability, while the growth of a rutile TeO$_2$ film on FeTe shows that epitaxy can also suppress it. 

Future work should extend this approach beyond the rutile $AX_2$ family and determine the finite-temperature stability of the predicted parent and daughter phases. 
It would also be interesting to investigate whether epitaxial strain, pressure, alloying, or circularly polarized fields can select one handedness and drive reversible chirality switching.

\section*{Acknowledgments}
We thanks M. Fava, E. E. McCabe, A. Romero and F. Gomez-Ortiz for fruitful discussions.
E.B. and S.M.T acknowledge the Fonds de la Recherche Scientifique (FNRS) PDR project CHRYSALID No.40003544. S.M.T.  acknowledges the Fonds de la Recherche Scientifique (FNRS) for financial support with a FRIA doctoral grant. S.A. acknowledge the Erasmus+ Mobility of  the Universit\'e de Li\`ege/Belgium. Calculations were carried out used the facilities provided by the CECI supercomputing center~\cite{ceci} thanks to the Universit\'e de Li\`ege/Belgium. 

%\bibliographystyle{apsrev4-2}
%\bibliography{References/reference, References/framework}

\begin{thebibliography}{55}%
\makeatletter
\providecommand \@ifxundefined [1]{%
 \@ifx{#1\undefined}
}%
\providecommand \@ifnum [1]{%
 \ifnum #1\expandafter \@firstoftwo
 \else \expandafter \@secondoftwo
 \fi
}%
\providecommand \@ifx [1]{%
 \ifx #1\expandafter \@firstoftwo
 \else \expandafter \@secondoftwo
 \fi
}%
\providecommand \natexlab [1]{#1}%
\providecommand \enquote  [1]{``#1''}%
\providecommand \bibnamefont  [1]{#1}%
\providecommand \bibfnamefont [1]{#1}%
\providecommand \citenamefont [1]{#1}%
\providecommand \href@noop [0]{\@secondoftwo}%
\providecommand \href [0]{\begingroup \@sanitize@url \@href}%
\providecommand \@href[1]{\@@startlink{#1}\@@href}%
\providecommand \@@href[1]{\endgroup#1\@@endlink}%
\providecommand \@sanitize@url [0]{\catcode `\\12\catcode `\$12\catcode
  `\&12\catcode `\#12\catcode `\^12\catcode `\_12\catcode `\%12\relax}%
\providecommand \@@startlink[1]{}%
\providecommand \@@endlink[0]{}%
\providecommand \url  [0]{\begingroup\@sanitize@url \@url }%
\providecommand \@url [1]{\endgroup\@href {#1}{\urlprefix }}%
\providecommand \urlprefix  [0]{URL }%
\providecommand \Eprint [0]{\href }%
\providecommand \doibase [0]{https://doi.org/}%
\providecommand \selectlanguage [0]{\@gobble}%
\providecommand \bibinfo  [0]{\@secondoftwo}%
\providecommand \bibfield  [0]{\@secondoftwo}%
\providecommand \translation [1]{[#1]}%
\providecommand \BibitemOpen [0]{}%
\providecommand \bibitemStop [0]{}%
\providecommand \bibitemNoStop [0]{.\EOS\space}%
\providecommand \EOS [0]{\spacefactor3000\relax}%
\providecommand \BibitemShut  [1]{\csname bibitem#1\endcsname}%
\let\auto@bib@innerbib\@empty
%</preamble>
\bibitem [{\citenamefont {Bousquet}\ \emph {et~al.}(2025)\citenamefont
  {Bousquet}, \citenamefont {Fava}, \citenamefont {Romestan}, \citenamefont
  {G{\'o}mez~Ortiz}, \citenamefont {McCabe},\ and\ \citenamefont
  {Romero}}]{Bousquet2025chirality}%
  \BibitemOpen
  \bibfield  {author} {\bibinfo {author} {\bibfnamefont {E.}~\bibnamefont
  {Bousquet}}, \bibinfo {author} {\bibfnamefont {M.}~\bibnamefont {Fava}},
  \bibinfo {author} {\bibfnamefont {Z.}~\bibnamefont {Romestan}}, \bibinfo
  {author} {\bibfnamefont {F.}~\bibnamefont {G{\'o}mez~Ortiz}}, \bibinfo
  {author} {\bibfnamefont {E.~E.}\ \bibnamefont {McCabe}},\ and\ \bibinfo
  {author} {\bibfnamefont {A.~H.}\ \bibnamefont {Romero}},\ }\href
  {https://doi.org/10.1088/1361-648X/adb674} {\bibfield  {journal} {\bibinfo
  {journal} {J. Phys.: Condens. Matter}\ }\textbf {\bibinfo {volume} {37}},\
  \bibinfo {pages} {163004} (\bibinfo {year} {2025})}\BibitemShut {NoStop}%
\bibitem [{\citenamefont {Juraschek}\ \emph {et~al.}(2025)\citenamefont
  {Juraschek}, \citenamefont {Geilhufe}, \citenamefont {Zhu} \emph
  {et~al.}}]{Juraschek2025chiralphonon}%
  \BibitemOpen
  \bibfield  {author} {\bibinfo {author} {\bibfnamefont {D.~M.}\ \bibnamefont
  {Juraschek}}, \bibinfo {author} {\bibfnamefont {R.~M.}\ \bibnamefont
  {Geilhufe}}, \bibinfo {author} {\bibfnamefont {H.}~\bibnamefont {Zhu}}, \emph
  {et~al.},\ }\href {https://doi.org/10.1038/s41567-025-03001-9} {\bibfield
  {journal} {\bibinfo  {journal} {Nature Physics}\ }\textbf {\bibinfo {volume}
  {21}},\ \bibinfo {pages} {1532} (\bibinfo {year} {2025})}\BibitemShut
  {NoStop}%
\bibitem [{\citenamefont {Ray}\ \emph {et~al.}(1999)\citenamefont {Ray},
  \citenamefont {Ananthavel}, \citenamefont {Waldeck},\ and\ \citenamefont
  {Naaman}}]{Ray1999CISS}%
  \BibitemOpen
  \bibfield  {author} {\bibinfo {author} {\bibfnamefont {K.}~\bibnamefont
  {Ray}}, \bibinfo {author} {\bibfnamefont {S.~P.}\ \bibnamefont {Ananthavel}},
  \bibinfo {author} {\bibfnamefont {D.~H.}\ \bibnamefont {Waldeck}},\ and\
  \bibinfo {author} {\bibfnamefont {R.}~\bibnamefont {Naaman}},\ }\href
  {https://doi.org/10.1126/science.283.5403.814} {\bibfield  {journal}
  {\bibinfo  {journal} {Science}\ }\textbf {\bibinfo {volume} {283}},\ \bibinfo
  {pages} {814} (\bibinfo {year} {1999})}\BibitemShut {NoStop}%
\bibitem [{\citenamefont {M{\"u}hlbauer}\ \emph {et~al.}(2009)\citenamefont
  {M{\"u}hlbauer}, \citenamefont {Binz}, \citenamefont {Jonietz}, \citenamefont
  {Pfleiderer}, \citenamefont {Rosch}, \citenamefont {Neubauer}, \citenamefont
  {Georgii},\ and\ \citenamefont {B{\"o}ni}}]{Muhlbauer2009skyrmion}%
  \BibitemOpen
  \bibfield  {author} {\bibinfo {author} {\bibfnamefont {S.}~\bibnamefont
  {M{\"u}hlbauer}}, \bibinfo {author} {\bibfnamefont {B.}~\bibnamefont {Binz}},
  \bibinfo {author} {\bibfnamefont {F.}~\bibnamefont {Jonietz}}, \bibinfo
  {author} {\bibfnamefont {C.}~\bibnamefont {Pfleiderer}}, \bibinfo {author}
  {\bibfnamefont {A.}~\bibnamefont {Rosch}}, \bibinfo {author} {\bibfnamefont
  {A.}~\bibnamefont {Neubauer}}, \bibinfo {author} {\bibfnamefont
  {R.}~\bibnamefont {Georgii}},\ and\ \bibinfo {author} {\bibfnamefont
  {P.}~\bibnamefont {B{\"o}ni}},\ }\href
  {https://doi.org/10.1126/science.1166767} {\bibfield  {journal} {\bibinfo
  {journal} {Science}\ }\textbf {\bibinfo {volume} {323}},\ \bibinfo {pages}
  {915} (\bibinfo {year} {2009})}\BibitemShut {NoStop}%
\bibitem [{\citenamefont {Bradlyn}\ \emph {et~al.}(2016)\citenamefont
  {Bradlyn}, \citenamefont {Cano}, \citenamefont {Wang}, \citenamefont
  {Vergniory}, \citenamefont {Felser}, \citenamefont {Cava},\ and\
  \citenamefont {Bernevig}}]{Bradlyn2016weylfermions}%
  \BibitemOpen
  \bibfield  {author} {\bibinfo {author} {\bibfnamefont {B.}~\bibnamefont
  {Bradlyn}}, \bibinfo {author} {\bibfnamefont {J.}~\bibnamefont {Cano}},
  \bibinfo {author} {\bibfnamefont {Z.}~\bibnamefont {Wang}}, \bibinfo {author}
  {\bibfnamefont {M.~G.}\ \bibnamefont {Vergniory}}, \bibinfo {author}
  {\bibfnamefont {C.}~\bibnamefont {Felser}}, \bibinfo {author} {\bibfnamefont
  {R.~J.}\ \bibnamefont {Cava}},\ and\ \bibinfo {author} {\bibfnamefont
  {B.~A.}\ \bibnamefont {Bernevig}},\ }\href
  {https://doi.org/10.1126/science.aaf5037} {\bibfield  {journal} {\bibinfo
  {journal} {Science}\ }\textbf {\bibinfo {volume} {353}},\ \bibinfo {pages}
  {aaf5037} (\bibinfo {year} {2016})}\BibitemShut {NoStop}%
\bibitem [{\citenamefont {Hasan}\ \emph {et~al.}(2021)\citenamefont {Hasan},
  \citenamefont {Chang}, \citenamefont {Belopolski}, \citenamefont {Bian},
  \citenamefont {Xu},\ and\ \citenamefont {Yin}}]{Hasan2021weylfermions}%
  \BibitemOpen
  \bibfield  {author} {\bibinfo {author} {\bibfnamefont {M.~Z.}\ \bibnamefont
  {Hasan}}, \bibinfo {author} {\bibfnamefont {G.}~\bibnamefont {Chang}},
  \bibinfo {author} {\bibfnamefont {I.}~\bibnamefont {Belopolski}}, \bibinfo
  {author} {\bibfnamefont {G.}~\bibnamefont {Bian}}, \bibinfo {author}
  {\bibfnamefont {S.-Y.}\ \bibnamefont {Xu}},\ and\ \bibinfo {author}
  {\bibfnamefont {J.-X.}\ \bibnamefont {Yin}},\ }\href
  {https://doi.org/10.1038/s41578-021-00345-0} {\bibfield  {journal} {\bibinfo
  {journal} {Nature Reviews Materials}\ }\textbf {\bibinfo {volume} {6}},\
  \bibinfo {pages} {784} (\bibinfo {year} {2021})}\BibitemShut {NoStop}%
\bibitem [{\citenamefont {\v{S}mejkal}\ \emph {et~al.}(2022)\citenamefont
  {\v{S}mejkal}, \citenamefont {Sinova},\ and\ \citenamefont
  {Jungwirth}}]{Smejkal2022altermagnetism}%
  \BibitemOpen
  \bibfield  {author} {\bibinfo {author} {\bibfnamefont {L.}~\bibnamefont
  {\v{S}mejkal}}, \bibinfo {author} {\bibfnamefont {J.}~\bibnamefont
  {Sinova}},\ and\ \bibinfo {author} {\bibfnamefont {T.}~\bibnamefont
  {Jungwirth}},\ }\href {https://doi.org/10.1103/PhysRevX.12.040501} {\bibfield
   {journal} {\bibinfo  {journal} {Physical Review X}\ }\textbf {\bibinfo
  {volume} {12}},\ \bibinfo {pages} {040501} (\bibinfo {year}
  {2022})}\BibitemShut {NoStop}%
\bibitem [{\citenamefont {\v{S}mejkal}\ \emph {et~al.}(2020)\citenamefont
  {\v{S}mejkal}, \citenamefont {Gonz{\'a}lez-Hern{\'a}ndez}, \citenamefont
  {Jungwirth},\ and\ \citenamefont {Sinova}}]{Smejkal2020halleffect}%
  \BibitemOpen
  \bibfield  {author} {\bibinfo {author} {\bibfnamefont {L.}~\bibnamefont
  {\v{S}mejkal}}, \bibinfo {author} {\bibfnamefont {R.}~\bibnamefont
  {Gonz{\'a}lez-Hern{\'a}ndez}}, \bibinfo {author} {\bibfnamefont
  {T.}~\bibnamefont {Jungwirth}},\ and\ \bibinfo {author} {\bibfnamefont
  {J.}~\bibnamefont {Sinova}},\ }\href {https://doi.org/10.1126/sciadv.aaz8809}
  {\bibfield  {journal} {\bibinfo  {journal} {Science Advances}\ }\textbf
  {\bibinfo {volume} {6}},\ \bibinfo {pages} {eaaz8809} (\bibinfo {year}
  {2020})}\BibitemShut {NoStop}%
\bibitem [{\citenamefont {Flack}(2003)}]{Flack2003chiralstructures}%
  \BibitemOpen
  \bibfield  {author} {\bibinfo {author} {\bibfnamefont {H.~D.}\ \bibnamefont
  {Flack}},\ }\href {https://doi.org/10.1002/hlca.200390109} {\bibfield
  {journal} {\bibinfo  {journal} {Helvetica Chimica Acta}\ }\textbf {\bibinfo
  {volume} {86}},\ \bibinfo {pages} {905} (\bibinfo {year} {2003})}\BibitemShut
  {NoStop}%
\bibitem [{\citenamefont {Fava}\ \emph
  {et~al.}(2025{\natexlab{a}})\citenamefont {Fava}, \citenamefont {McCabe},
  \citenamefont {Romero},\ and\ \citenamefont {Bousquet}}]{Fava2025k3nio2}%
  \BibitemOpen
  \bibfield  {author} {\bibinfo {author} {\bibfnamefont {M.}~\bibnamefont
  {Fava}}, \bibinfo {author} {\bibfnamefont {E.}~\bibnamefont {McCabe}},
  \bibinfo {author} {\bibfnamefont {A.~H.}\ \bibnamefont {Romero}},\ and\
  \bibinfo {author} {\bibfnamefont {E.}~\bibnamefont {Bousquet}},\ }\href
  {https://doi.org/10.1103/PhysRevB.111.174102} {\bibfield  {journal} {\bibinfo
   {journal} {Phys. Rev. B}\ }\textbf {\bibinfo {volume} {111}},\ \bibinfo
  {pages} {174102} (\bibinfo {year} {2025}{\natexlab{a}})}\BibitemShut
  {NoStop}%
\bibitem [{\citenamefont {G{\'o}mez-Ortiz}\ \emph {et~al.}(2025)\citenamefont
  {G{\'o}mez-Ortiz}, \citenamefont {Romero},\ and\ \citenamefont
  {Bousquet}}]{GomezOrtiz2025pathways}%
  \BibitemOpen
  \bibfield  {author} {\bibinfo {author} {\bibfnamefont {F.}~\bibnamefont
  {G{\'o}mez-Ortiz}}, \bibinfo {author} {\bibfnamefont {A.~H.}\ \bibnamefont
  {Romero}},\ and\ \bibinfo {author} {\bibfnamefont {E.}~\bibnamefont
  {Bousquet}},\ }\href {https://doi.org/10.1103/PhysRevB.112.014107} {\bibfield
   {journal} {\bibinfo  {journal} {Phys. Rev. B}\ }\textbf {\bibinfo {volume}
  {112}},\ \bibinfo {pages} {014107} (\bibinfo {year} {2025})}\BibitemShut
  {NoStop}%
\bibitem [{\citenamefont {Fava}\ \emph
  {et~al.}(2025{\natexlab{b}})\citenamefont {Fava}, \citenamefont {Romero},\
  and\ \citenamefont {Bousquet}}]{Fava2025handedness}%
  \BibitemOpen
  \bibfield  {author} {\bibinfo {author} {\bibfnamefont {M.}~\bibnamefont
  {Fava}}, \bibinfo {author} {\bibfnamefont {A.~H.}\ \bibnamefont {Romero}},\
  and\ \bibinfo {author} {\bibfnamefont {E.}~\bibnamefont {Bousquet}},\ }\href
  {https://doi.org/10.1103/c4v9-nd4b} {\bibfield  {journal} {\bibinfo
  {journal} {Phys. Rev. Lett.}\ }\textbf {\bibinfo {volume} {135}},\ \bibinfo
  {pages} {146102} (\bibinfo {year} {2025}{\natexlab{b}})}\BibitemShut
  {NoStop}%
\bibitem [{\citenamefont {Spaldin}(2026)}]{spaldin2026}%
  \BibitemOpen
  \bibfield  {author} {\bibinfo {author} {\bibfnamefont {N.~A.}\ \bibnamefont
  {Spaldin}},\ }\href
  {https://doi.org/https://doi.org/10.1016/j.newton.2026.100581} {\bibfield
  {journal} {\bibinfo  {journal} {Newton}\ ,\ \bibinfo {pages} {100581}}
  (\bibinfo {year} {2026})}\BibitemShut {NoStop}%
\bibitem [{\citenamefont {G{\'o}mez-Ortiz}\ \emph {et~al.}(2026)\citenamefont
  {G{\'o}mez-Ortiz}, \citenamefont {Mamoudou~Taganga}, \citenamefont {McCabe},
  \citenamefont {Romero},\ and\ \citenamefont
  {Bousquet}}]{GomezOrtiz2026ferroic}%
  \BibitemOpen
  \bibfield  {author} {\bibinfo {author} {\bibfnamefont {F.}~\bibnamefont
  {G{\'o}mez-Ortiz}}, \bibinfo {author} {\bibfnamefont {S.}~\bibnamefont
  {Mamoudou~Taganga}}, \bibinfo {author} {\bibfnamefont {E.~E.}\ \bibnamefont
  {McCabe}}, \bibinfo {author} {\bibfnamefont {A.~H.}\ \bibnamefont {Romero}},\
  and\ \bibinfo {author} {\bibfnamefont {E.}~\bibnamefont {Bousquet}},\ }\href
  {https://doi.org/10.1103/ktz2-7yd5} {\bibfield  {journal} {\bibinfo
  {journal} {Phys. Rev. Lett.}\ }\textbf {\bibinfo {volume} {137}},\ \bibinfo
  {pages} {036103} (\bibinfo {year} {2026})}\BibitemShut {NoStop}%
\bibitem [{\citenamefont {Baur}(2007)}]{Baur01012007rutile}%
  \BibitemOpen
  \bibfield  {author} {\bibinfo {author} {\bibfnamefont {W.~H.}\ \bibnamefont
  {Baur}},\ }\href {https://doi.org/10.1080/08893110701433435} {\bibfield
  {journal} {\bibinfo  {journal} {Crystallography Reviews}\ }\textbf {\bibinfo
  {volume} {13}},\ \bibinfo {pages} {65} (\bibinfo {year} {2007})},\ \Eprint
  {https://arxiv.org/abs/https://www.tandfonline.com/doi/pdf/10.1080/08893110701433435}
  {https://www.tandfonline.com/doi/pdf/10.1080/08893110701433435} \BibitemShut
  {NoStop}%
\bibitem [{\citenamefont {Den{\`e}s}\ \emph {et~al.}(1979)\citenamefont
  {Den{\`e}s}, \citenamefont {Pannetier}, \citenamefont {Lucas},\ and\
  \citenamefont {Le~Marouille}}]{Denes1979alpha}%
  \BibitemOpen
  \bibfield  {author} {\bibinfo {author} {\bibfnamefont {G.}~\bibnamefont
  {Den{\`e}s}}, \bibinfo {author} {\bibfnamefont {J.}~\bibnamefont
  {Pannetier}}, \bibinfo {author} {\bibfnamefont {J.}~\bibnamefont {Lucas}},\
  and\ \bibinfo {author} {\bibfnamefont {J.~Y.}\ \bibnamefont {Le~Marouille}},\
  }\href {https://doi.org/10.1016/0022-4596(79)90245-7} {\bibfield  {journal}
  {\bibinfo  {journal} {J. Solid State Chem.}\ }\textbf {\bibinfo {volume}
  {30}},\ \bibinfo {pages} {335} (\bibinfo {year} {1979})}\BibitemShut
  {NoStop}%
\bibitem [{\citenamefont {Den{\`e}s}\ \emph {et~al.}(1980)\citenamefont
  {Den{\`e}s}, \citenamefont {Pannetier},\ and\ \citenamefont
  {Lucas}}]{Denes1980structure}%
  \BibitemOpen
  \bibfield  {author} {\bibinfo {author} {\bibfnamefont {G.}~\bibnamefont
  {Den{\`e}s}}, \bibinfo {author} {\bibfnamefont {J.}~\bibnamefont
  {Pannetier}},\ and\ \bibinfo {author} {\bibfnamefont {J.}~\bibnamefont
  {Lucas}},\ }\href {https://doi.org/10.1016/0022-4596(80)90541-1} {\bibfield
  {journal} {\bibinfo  {journal} {J. Solid State Chem.}\ }\textbf {\bibinfo
  {volume} {33}},\ \bibinfo {pages} {1} (\bibinfo {year} {1980})}\BibitemShut
  {NoStop}%
\bibitem [{\citenamefont {Den{\`e}s}(1980)}]{Denes1980transitions}%
  \BibitemOpen
  \bibfield  {author} {\bibinfo {author} {\bibfnamefont {G.}~\bibnamefont
  {Den{\`e}s}},\ }\href {https://doi.org/10.1016/0025-5408(80)90015-X}
  {\bibfield  {journal} {\bibinfo  {journal} {Mater. Res. Bull.}\ }\textbf
  {\bibinfo {volume} {15}},\ \bibinfo {pages} {807} (\bibinfo {year}
  {1980})}\BibitemShut {NoStop}%
\bibitem [{\citenamefont {Dénès}(1981)}]{DENES198120thermalexpansion}%
  \BibitemOpen
  \bibfield  {author} {\bibinfo {author} {\bibfnamefont {G.}~\bibnamefont
  {Dénès}},\ }\href
  {https://doi.org/https://doi.org/10.1016/0022-4596(81)90187-0} {\bibfield
  {journal} {\bibinfo  {journal} {Journal of Solid State Chemistry}\ }\textbf
  {\bibinfo {volume} {36}},\ \bibinfo {pages} {20} (\bibinfo {year}
  {1981})}\BibitemShut {NoStop}%
\bibitem [{\citenamefont {Den{\`e}s}(1981)}]{Denes1981kinetics}%
  \BibitemOpen
  \bibfield  {author} {\bibinfo {author} {\bibfnamefont {G.}~\bibnamefont
  {Den{\`e}s}},\ }\href {https://doi.org/10.1016/0022-4596(81)90003-7}
  {\bibfield  {journal} {\bibinfo  {journal} {J. Solid State Chem.}\ }\textbf
  {\bibinfo {volume} {37}},\ \bibinfo {pages} {16} (\bibinfo {year}
  {1981})}\BibitemShut {NoStop}%
\bibitem [{\citenamefont {Thomas}(1988)}]{pathomas1988}%
  \BibitemOpen
  \bibfield  {author} {\bibinfo {author} {\bibfnamefont {P.~A.}\ \bibnamefont
  {Thomas}},\ }\href {https://doi.org/10.1088/0022-3719/21/25/009} {\bibfield
  {journal} {\bibinfo  {journal} {J. Phys. C: Solid State Phys.}\ }\textbf
  {\bibinfo {volume} {21}},\ \bibinfo {pages} {4611} (\bibinfo {year}
  {1988})}\BibitemShut {NoStop}%
\bibitem [{\citenamefont {Peng}\ \emph {et~al.}(2022)\citenamefont {Peng},
  \citenamefont {Ren}, \citenamefont {Wu}, \citenamefont {Ru}, \citenamefont
  {Lu}, \citenamefont {Chen}, \citenamefont {Wang}, \citenamefont {Fang},
  \citenamefont {Li}, \citenamefont {Chi}, \citenamefont {Ding}, \citenamefont
  {Wang}, \citenamefont {Wang},\ and\ \citenamefont {Li}}]{Peng2022teo2}%
  \BibitemOpen
  \bibfield  {author} {\bibinfo {author} {\bibfnamefont {K.}~\bibnamefont
  {Peng}}, \bibinfo {author} {\bibfnamefont {W.}~\bibnamefont {Ren}}, \bibinfo
  {author} {\bibfnamefont {Y.}~\bibnamefont {Wu}}, \bibinfo {author}
  {\bibfnamefont {H.}~\bibnamefont {Ru}}, \bibinfo {author} {\bibfnamefont
  {S.}~\bibnamefont {Lu}}, \bibinfo {author} {\bibfnamefont {A.}~\bibnamefont
  {Chen}}, \bibinfo {author} {\bibfnamefont {P.}~\bibnamefont {Wang}}, \bibinfo
  {author} {\bibfnamefont {X.}~\bibnamefont {Fang}}, \bibinfo {author}
  {\bibfnamefont {H.}~\bibnamefont {Li}}, \bibinfo {author} {\bibfnamefont
  {L.}~\bibnamefont {Chi}}, \bibinfo {author} {\bibfnamefont {S.}~\bibnamefont
  {Ding}}, \bibinfo {author} {\bibfnamefont {L.}~\bibnamefont {Wang}}, \bibinfo
  {author} {\bibfnamefont {Y.}~\bibnamefont {Wang}},\ and\ \bibinfo {author}
  {\bibfnamefont {F.}~\bibnamefont {Li}},\ }\href
  {https://doi.org/10.1007/s10853-022-07278-6} {\bibfield  {journal} {\bibinfo
  {journal} {J. Mater. Sci.}\ }\textbf {\bibinfo {volume} {57}},\ \bibinfo
  {pages} {10225} (\bibinfo {year} {2022})}\BibitemShut {NoStop}%
\bibitem [{\citenamefont {Gonze}\ \emph {et~al.}(2016)\citenamefont {Gonze},
  \citenamefont {Jollet}, \citenamefont {Abreu~Araujo}, \citenamefont {Adams},
  \citenamefont {Amadon}, \citenamefont {Applencourt}, \citenamefont {Audouze},
  \citenamefont {Beuken}, \citenamefont {Bieder}, \citenamefont {Bokhanchuk},
  \citenamefont {Bousquet}, \citenamefont {Bruneval}, \citenamefont {Caliste},
  \citenamefont {Cote}, \citenamefont {Dahm}, \citenamefont {Da~Pieve},
  \citenamefont {Delaveau}, \citenamefont {Di~Gennaro}, \citenamefont {Dorado},
  \citenamefont {Espejo}, \citenamefont {Geneste}, \citenamefont {Genovese},
  \citenamefont {Gerossier}, \citenamefont {Giantomassi}, \citenamefont
  {Gillet}, \citenamefont {Hamann}, \citenamefont {He}, \citenamefont {Jomard},
  \citenamefont {Laflamme~Janssen}, \citenamefont {Le~Roux}, \citenamefont
  {Levitt}, \citenamefont {Lherbier}, \citenamefont {Liu}, \citenamefont
  {Lukacevic}, \citenamefont {Martin}, \citenamefont {Martins}, \citenamefont
  {Oliveira}, \citenamefont {Ponce}, \citenamefont {Pouillon}, \citenamefont
  {Rangel}, \citenamefont {Rignanese}, \citenamefont {Romero}, \citenamefont
  {Rousseau}, \citenamefont {Rubel}, \citenamefont {Shukri}, \citenamefont
  {Stankovski}, \citenamefont {Torrent}, \citenamefont {Van~Setten},
  \citenamefont {Van~Troeye}, \citenamefont {Verstraete}, \citenamefont
  {Waroquier}, \citenamefont {Wiktor}, \citenamefont {Xu}, \citenamefont
  {Zhou},\ and\ \citenamefont {Zwanziger}}]{abinit1}%
  \BibitemOpen
  \bibfield  {author} {\bibinfo {author} {\bibfnamefont {X.}~\bibnamefont
  {Gonze}}, \bibinfo {author} {\bibfnamefont {F.}~\bibnamefont {Jollet}},
  \bibinfo {author} {\bibfnamefont {F.}~\bibnamefont {Abreu~Araujo}}, \bibinfo
  {author} {\bibfnamefont {D.}~\bibnamefont {Adams}}, \bibinfo {author}
  {\bibfnamefont {B.}~\bibnamefont {Amadon}}, \bibinfo {author} {\bibfnamefont
  {T.}~\bibnamefont {Applencourt}}, \bibinfo {author} {\bibfnamefont
  {C.}~\bibnamefont {Audouze}}, \bibinfo {author} {\bibfnamefont {J.-M.}\
  \bibnamefont {Beuken}}, \bibinfo {author} {\bibfnamefont {J.}~\bibnamefont
  {Bieder}}, \bibinfo {author} {\bibfnamefont {A.}~\bibnamefont {Bokhanchuk}},
  \bibinfo {author} {\bibfnamefont {E.}~\bibnamefont {Bousquet}}, \bibinfo
  {author} {\bibfnamefont {F.}~\bibnamefont {Bruneval}}, \bibinfo {author}
  {\bibfnamefont {D.}~\bibnamefont {Caliste}}, \bibinfo {author} {\bibfnamefont
  {M.}~\bibnamefont {Cote}}, \bibinfo {author} {\bibfnamefont {F.}~\bibnamefont
  {Dahm}}, \bibinfo {author} {\bibfnamefont {F.}~\bibnamefont {Da~Pieve}},
  \bibinfo {author} {\bibfnamefont {M.}~\bibnamefont {Delaveau}}, \bibinfo
  {author} {\bibfnamefont {M.}~\bibnamefont {Di~Gennaro}}, \bibinfo {author}
  {\bibfnamefont {B.}~\bibnamefont {Dorado}}, \bibinfo {author} {\bibfnamefont
  {C.}~\bibnamefont {Espejo}}, \bibinfo {author} {\bibfnamefont
  {G.}~\bibnamefont {Geneste}}, \bibinfo {author} {\bibfnamefont
  {L.}~\bibnamefont {Genovese}}, \bibinfo {author} {\bibfnamefont
  {A.}~\bibnamefont {Gerossier}}, \bibinfo {author} {\bibfnamefont
  {M.}~\bibnamefont {Giantomassi}}, \bibinfo {author} {\bibfnamefont
  {Y.}~\bibnamefont {Gillet}}, \bibinfo {author} {\bibfnamefont {D.~R.}\
  \bibnamefont {Hamann}}, \bibinfo {author} {\bibfnamefont {L.}~\bibnamefont
  {He}}, \bibinfo {author} {\bibfnamefont {G.}~\bibnamefont {Jomard}}, \bibinfo
  {author} {\bibfnamefont {J.}~\bibnamefont {Laflamme~Janssen}}, \bibinfo
  {author} {\bibfnamefont {S.}~\bibnamefont {Le~Roux}}, \bibinfo {author}
  {\bibfnamefont {A.}~\bibnamefont {Levitt}}, \bibinfo {author} {\bibfnamefont
  {A.}~\bibnamefont {Lherbier}}, \bibinfo {author} {\bibfnamefont
  {F.}~\bibnamefont {Liu}}, \bibinfo {author} {\bibfnamefont {I.}~\bibnamefont
  {Lukacevic}}, \bibinfo {author} {\bibfnamefont {A.}~\bibnamefont {Martin}},
  \bibinfo {author} {\bibfnamefont {C.}~\bibnamefont {Martins}}, \bibinfo
  {author} {\bibfnamefont {M.~J.~T.}\ \bibnamefont {Oliveira}}, \bibinfo
  {author} {\bibfnamefont {S.}~\bibnamefont {Ponce}}, \bibinfo {author}
  {\bibfnamefont {Y.}~\bibnamefont {Pouillon}}, \bibinfo {author}
  {\bibfnamefont {T.}~\bibnamefont {Rangel}}, \bibinfo {author} {\bibfnamefont
  {G.-M.}\ \bibnamefont {Rignanese}}, \bibinfo {author} {\bibfnamefont {A.~H.}\
  \bibnamefont {Romero}}, \bibinfo {author} {\bibfnamefont {B.}~\bibnamefont
  {Rousseau}}, \bibinfo {author} {\bibfnamefont {O.}~\bibnamefont {Rubel}},
  \bibinfo {author} {\bibfnamefont {A.~A.}\ \bibnamefont {Shukri}}, \bibinfo
  {author} {\bibfnamefont {M.}~\bibnamefont {Stankovski}}, \bibinfo {author}
  {\bibfnamefont {M.}~\bibnamefont {Torrent}}, \bibinfo {author} {\bibfnamefont
  {M.~J.}\ \bibnamefont {Van~Setten}}, \bibinfo {author} {\bibfnamefont
  {B.}~\bibnamefont {Van~Troeye}}, \bibinfo {author} {\bibfnamefont {M.~J.}\
  \bibnamefont {Verstraete}}, \bibinfo {author} {\bibfnamefont
  {D.}~\bibnamefont {Waroquier}}, \bibinfo {author} {\bibfnamefont
  {J.}~\bibnamefont {Wiktor}}, \bibinfo {author} {\bibfnamefont
  {B.}~\bibnamefont {Xu}}, \bibinfo {author} {\bibfnamefont {A.}~\bibnamefont
  {Zhou}},\ and\ \bibinfo {author} {\bibfnamefont {J.~W.}\ \bibnamefont
  {Zwanziger}},\ }\href
  {https://www.abinit.org/sites/default/files/ABINIT16.pdf} {\bibfield
  {journal} {\bibinfo  {journal} {Computer Physics Communications}\ }\textbf
  {\bibinfo {volume} {205}},\ \bibinfo {pages} {106} (\bibinfo {year}
  {2016})}\BibitemShut {NoStop}%
\bibitem [{\citenamefont {Gonze}\ \emph {et~al.}(2020)\citenamefont {Gonze},
  \citenamefont {Amadon}, \citenamefont {Antonius}, \citenamefont {Arnardi},
  \citenamefont {Baguet}, \citenamefont {Beuken}, \citenamefont {Bieder},
  \citenamefont {Bottin}, \citenamefont {Bouchet}, \citenamefont {Bousquet},
  \citenamefont {Brouwer}, \citenamefont {Bruneval}, \citenamefont {Brunin},
  \citenamefont {Cavignac}, \citenamefont {Charraud}, \citenamefont {Chen},
  \citenamefont {Cote}, \citenamefont {Cottenier}, \citenamefont {Denier},
  \citenamefont {Geneste}, \citenamefont {Ghosez}, \citenamefont {Giantomassi},
  \citenamefont {Gillet}, \citenamefont {Gingras}, \citenamefont {Hamann},
  \citenamefont {Hautier}, \citenamefont {He}, \citenamefont {Helbig},
  \citenamefont {Holzwarth}, \citenamefont {Jia}, \citenamefont {Jollet},
  \citenamefont {Lafargue-Dit-Hauret}, \citenamefont {Lejaeghere},
  \citenamefont {Marques}, \citenamefont {Martin}, \citenamefont {Martins},
  \citenamefont {Miranda}, \citenamefont {Naccarato}, \citenamefont {Persson},
  \citenamefont {Petretto}, \citenamefont {Planes}, \citenamefont {Pouillon},
  \citenamefont {Prokhorenko}, \citenamefont {Ricci}, \citenamefont
  {Rignanese}, \citenamefont {Romero}, \citenamefont {Schmitt}, \citenamefont
  {Torrent}, \citenamefont {van Setten}, \citenamefont {Van~Troeye},
  \citenamefont {Verstraete}, \citenamefont {Zerah},\ and\ \citenamefont
  {Zwanziger}}]{abinit2}%
  \BibitemOpen
  \bibfield  {author} {\bibinfo {author} {\bibfnamefont {X.}~\bibnamefont
  {Gonze}}, \bibinfo {author} {\bibfnamefont {B.}~\bibnamefont {Amadon}},
  \bibinfo {author} {\bibfnamefont {G.}~\bibnamefont {Antonius}}, \bibinfo
  {author} {\bibfnamefont {F.}~\bibnamefont {Arnardi}}, \bibinfo {author}
  {\bibfnamefont {L.}~\bibnamefont {Baguet}}, \bibinfo {author} {\bibfnamefont
  {J.-M.}\ \bibnamefont {Beuken}}, \bibinfo {author} {\bibfnamefont
  {J.}~\bibnamefont {Bieder}}, \bibinfo {author} {\bibfnamefont
  {F.}~\bibnamefont {Bottin}}, \bibinfo {author} {\bibfnamefont
  {J.}~\bibnamefont {Bouchet}}, \bibinfo {author} {\bibfnamefont
  {E.}~\bibnamefont {Bousquet}}, \bibinfo {author} {\bibfnamefont
  {N.}~\bibnamefont {Brouwer}}, \bibinfo {author} {\bibfnamefont
  {F.}~\bibnamefont {Bruneval}}, \bibinfo {author} {\bibfnamefont
  {G.}~\bibnamefont {Brunin}}, \bibinfo {author} {\bibfnamefont
  {T.}~\bibnamefont {Cavignac}}, \bibinfo {author} {\bibfnamefont {J.-B.}\
  \bibnamefont {Charraud}}, \bibinfo {author} {\bibfnamefont {W.}~\bibnamefont
  {Chen}}, \bibinfo {author} {\bibfnamefont {M.}~\bibnamefont {Cote}}, \bibinfo
  {author} {\bibfnamefont {S.}~\bibnamefont {Cottenier}}, \bibinfo {author}
  {\bibfnamefont {J.}~\bibnamefont {Denier}}, \bibinfo {author} {\bibfnamefont
  {G.}~\bibnamefont {Geneste}}, \bibinfo {author} {\bibfnamefont
  {P.}~\bibnamefont {Ghosez}}, \bibinfo {author} {\bibfnamefont
  {M.}~\bibnamefont {Giantomassi}}, \bibinfo {author} {\bibfnamefont
  {Y.}~\bibnamefont {Gillet}}, \bibinfo {author} {\bibfnamefont
  {O.}~\bibnamefont {Gingras}}, \bibinfo {author} {\bibfnamefont {D.~R.}\
  \bibnamefont {Hamann}}, \bibinfo {author} {\bibfnamefont {G.}~\bibnamefont
  {Hautier}}, \bibinfo {author} {\bibfnamefont {X.}~\bibnamefont {He}},
  \bibinfo {author} {\bibfnamefont {N.}~\bibnamefont {Helbig}}, \bibinfo
  {author} {\bibfnamefont {N.}~\bibnamefont {Holzwarth}}, \bibinfo {author}
  {\bibfnamefont {Y.}~\bibnamefont {Jia}}, \bibinfo {author} {\bibfnamefont
  {F.}~\bibnamefont {Jollet}}, \bibinfo {author} {\bibfnamefont
  {W.}~\bibnamefont {Lafargue-Dit-Hauret}}, \bibinfo {author} {\bibfnamefont
  {K.}~\bibnamefont {Lejaeghere}}, \bibinfo {author} {\bibfnamefont {M.~A.~L.}\
  \bibnamefont {Marques}}, \bibinfo {author} {\bibfnamefont {A.}~\bibnamefont
  {Martin}}, \bibinfo {author} {\bibfnamefont {C.}~\bibnamefont {Martins}},
  \bibinfo {author} {\bibfnamefont {H.~P.~C.}\ \bibnamefont {Miranda}},
  \bibinfo {author} {\bibfnamefont {F.}~\bibnamefont {Naccarato}}, \bibinfo
  {author} {\bibfnamefont {K.}~\bibnamefont {Persson}}, \bibinfo {author}
  {\bibfnamefont {G.}~\bibnamefont {Petretto}}, \bibinfo {author}
  {\bibfnamefont {V.}~\bibnamefont {Planes}}, \bibinfo {author} {\bibfnamefont
  {Y.}~\bibnamefont {Pouillon}}, \bibinfo {author} {\bibfnamefont
  {S.}~\bibnamefont {Prokhorenko}}, \bibinfo {author} {\bibfnamefont
  {F.}~\bibnamefont {Ricci}}, \bibinfo {author} {\bibfnamefont {G.-M.}\
  \bibnamefont {Rignanese}}, \bibinfo {author} {\bibfnamefont {A.~H.}\
  \bibnamefont {Romero}}, \bibinfo {author} {\bibfnamefont {M.~M.}\
  \bibnamefont {Schmitt}}, \bibinfo {author} {\bibfnamefont {M.}~\bibnamefont
  {Torrent}}, \bibinfo {author} {\bibfnamefont {M.~J.}\ \bibnamefont {van
  Setten}}, \bibinfo {author} {\bibfnamefont {B.}~\bibnamefont {Van~Troeye}},
  \bibinfo {author} {\bibfnamefont {M.~J.}\ \bibnamefont {Verstraete}},
  \bibinfo {author} {\bibfnamefont {G.}~\bibnamefont {Zerah}},\ and\ \bibinfo
  {author} {\bibfnamefont {J.~W.}\ \bibnamefont {Zwanziger}},\ }\href
  {https://doi.org/10.1016/j.cpc.2019.107042} {\bibfield  {journal} {\bibinfo
  {journal} {Computer Physics Communications}\ }\textbf {\bibinfo {volume}
  {248}},\ \bibinfo {pages} {107042} (\bibinfo {year} {2020})}\BibitemShut
  {NoStop}%
\bibitem [{abi()}]{abinit-web}%
  \BibitemOpen
  \href {http://www.abinit.org} {\bibinfo {title} {Abinit
  software}}\BibitemShut {NoStop}%
\bibitem [{\citenamefont {Marques}\ \emph {et~al.}(2012)\citenamefont
  {Marques}, \citenamefont {Oliveira},\ and\ \citenamefont {Burnus}}]{marques}%
  \BibitemOpen
  \bibfield  {author} {\bibinfo {author} {\bibfnamefont {M.~A.~L.}\
  \bibnamefont {Marques}}, \bibinfo {author} {\bibfnamefont {M.~J.~T.}\
  \bibnamefont {Oliveira}},\ and\ \bibinfo {author} {\bibfnamefont
  {T.}~\bibnamefont {Burnus}},\ }\href
  {https://doi.org/10.1016/j.cpc.2012.05.007} {\bibfield  {journal} {\bibinfo
  {journal} {Computer Physics Communications}\ }\textbf {\bibinfo {volume}
  {183}},\ \bibinfo {pages} {2272} (\bibinfo {year} {2012})}\BibitemShut
  {NoStop}%
\bibitem [{\citenamefont {Hamann}(2013)}]{hamann}%
  \BibitemOpen
  \bibfield  {author} {\bibinfo {author} {\bibfnamefont {D.~R.}\ \bibnamefont
  {Hamann}},\ }\href {https://doi.org/10.1103/PhysRevB.88.085117} {\bibfield
  {journal} {\bibinfo  {journal} {Physical Review B}\ }\textbf {\bibinfo
  {volume} {88}},\ \bibinfo {pages} {085117} (\bibinfo {year}
  {2013})}\BibitemShut {NoStop}%
\bibitem [{\citenamefont {van Setten}\ \emph {et~al.}(2018)\citenamefont {van
  Setten}, \citenamefont {Giantomassi}, \citenamefont {Bousquet}, \citenamefont
  {Verstraete}, \citenamefont {Hamann}, \citenamefont {Gonze},\ and\
  \citenamefont {Rignanese}}]{Setten}%
  \BibitemOpen
  \bibfield  {author} {\bibinfo {author} {\bibfnamefont {M.~J.}\ \bibnamefont
  {van Setten}}, \bibinfo {author} {\bibfnamefont {M.}~\bibnamefont
  {Giantomassi}}, \bibinfo {author} {\bibfnamefont {E.}~\bibnamefont
  {Bousquet}}, \bibinfo {author} {\bibfnamefont {M.~J.}\ \bibnamefont
  {Verstraete}}, \bibinfo {author} {\bibfnamefont {D.~R.}\ \bibnamefont
  {Hamann}}, \bibinfo {author} {\bibfnamefont {X.}~\bibnamefont {Gonze}},\ and\
  \bibinfo {author} {\bibfnamefont {G.-M.}\ \bibnamefont {Rignanese}},\ }\href
  {https://doi.org/10.1016/j.cpc.2018.01.012} {\bibfield  {journal} {\bibinfo
  {journal} {Computer Physics Communications}\ }\textbf {\bibinfo {volume}
  {226}},\ \bibinfo {pages} {39} (\bibinfo {year} {2018})}\BibitemShut
  {NoStop}%
\bibitem [{pse()}]{pseudojo}%
  \BibitemOpen
  \href {http://www.pseudo-dojo.org/index.html} {\bibinfo {title}
  {Pseudodojo}}\BibitemShut {NoStop}%
\bibitem [{\citenamefont {Perdew}\ \emph {et~al.}(2008)\citenamefont {Perdew},
  \citenamefont {Ruzsinszky}, \citenamefont {Csonka}, \citenamefont {Vydrov},
  \citenamefont {Scuseria}, \citenamefont {Constantin}, \citenamefont {Zhou},\
  and\ \citenamefont {Burke}}]{perdew}%
  \BibitemOpen
  \bibfield  {author} {\bibinfo {author} {\bibfnamefont {J.~P.}\ \bibnamefont
  {Perdew}}, \bibinfo {author} {\bibfnamefont {A.}~\bibnamefont {Ruzsinszky}},
  \bibinfo {author} {\bibfnamefont {G.~I.}\ \bibnamefont {Csonka}}, \bibinfo
  {author} {\bibfnamefont {O.}~\bibnamefont {Vydrov}}, \bibinfo {author}
  {\bibfnamefont {G.}~\bibnamefont {Scuseria}}, \bibinfo {author}
  {\bibfnamefont {L.}~\bibnamefont {Constantin}}, \bibinfo {author}
  {\bibfnamefont {X.}~\bibnamefont {Zhou}},\ and\ \bibinfo {author}
  {\bibfnamefont {K.}~\bibnamefont {Burke}},\ }\href
  {https://doi.org/10.1103/PhysRevLett.100.136406} {\bibfield  {journal}
  {\bibinfo  {journal} {Physical Review Letters}\ }\textbf {\bibinfo {volume}
  {100}},\ \bibinfo {pages} {136406} (\bibinfo {year} {2008})}\BibitemShut
  {NoStop}%
\bibitem [{\citenamefont {Gonze}\ and\ \citenamefont {Lee}(1997)}]{gonze}%
  \BibitemOpen
  \bibfield  {author} {\bibinfo {author} {\bibfnamefont {X.}~\bibnamefont
  {Gonze}}\ and\ \bibinfo {author} {\bibfnamefont {C.}~\bibnamefont {Lee}},\
  }\href {https://doi.org/10.1103/PhysRevB.55.10355} {\bibfield  {journal}
  {\bibinfo  {journal} {Physical Review B}\ }\textbf {\bibinfo {volume} {55}},\
  \bibinfo {pages} {10355} (\bibinfo {year} {1997})}\BibitemShut {NoStop}%
\bibitem [{\citenamefont {Stokes}\ and\ \citenamefont {Hatch}(2005)}]{findsym}%
  \BibitemOpen
  \bibfield  {author} {\bibinfo {author} {\bibfnamefont {H.~T.}\ \bibnamefont
  {Stokes}}\ and\ \bibinfo {author} {\bibfnamefont {D.~M.}\ \bibnamefont
  {Hatch}},\ }\href {http://stokes.byu.edu/iso/findsym.php} {\bibfield
  {journal} {\bibinfo  {journal} {Journal of Applied Crystallography}\ }\textbf
  {\bibinfo {volume} {38}},\ \bibinfo {pages} {237} (\bibinfo {year}
  {2005})}\BibitemShut {NoStop}%
\bibitem [{\citenamefont {Stokes}\ \emph {et~al.}(2007)\citenamefont {Stokes},
  \citenamefont {Hatch},\ and\ \citenamefont {Campbell}}]{isotropycode}%
  \BibitemOpen
  \bibfield  {author} {\bibinfo {author} {\bibfnamefont {H.~T.}\ \bibnamefont
  {Stokes}}, \bibinfo {author} {\bibfnamefont {D.~M.}\ \bibnamefont {Hatch}},\
  and\ \bibinfo {author} {\bibfnamefont {B.~J.}\ \bibnamefont {Campbell}},\
  }\href {http://stokes.byu.edu/isotropy.html} {\bibinfo {title} {Isotropy
  software suite}} (\bibinfo {year} {2007})\BibitemShut {NoStop}%
\bibitem [{\citenamefont {Orobengoa}\ \emph {et~al.}(2009)\citenamefont
  {Orobengoa}, \citenamefont {Capillas}, \citenamefont {Aroyo},\ and\
  \citenamefont {Perez-Mato}}]{amplimodecode}%
  \BibitemOpen
  \bibfield  {author} {\bibinfo {author} {\bibfnamefont {D.}~\bibnamefont
  {Orobengoa}}, \bibinfo {author} {\bibfnamefont {C.}~\bibnamefont {Capillas}},
  \bibinfo {author} {\bibfnamefont {M.~I.}\ \bibnamefont {Aroyo}},\ and\
  \bibinfo {author} {\bibfnamefont {J.~M.}\ \bibnamefont {Perez-Mato}},\ }\href
  {http://www.cryst.ehu.es/cryst/amplimodes.html} {\bibfield  {journal}
  {\bibinfo  {journal} {Journal of Applied Crystallography}\ }\textbf {\bibinfo
  {volume} {42}},\ \bibinfo {pages} {820} (\bibinfo {year} {2009})}\BibitemShut
  {NoStop}%
\bibitem [{\citenamefont {Campbell}\ \emph {et~al.}(2006)\citenamefont
  {Campbell}, \citenamefont {Stokes}, \citenamefont {Tanner},\ and\
  \citenamefont {Hatch}}]{isodistortcode}%
  \BibitemOpen
  \bibfield  {author} {\bibinfo {author} {\bibfnamefont {B.~J.}\ \bibnamefont
  {Campbell}}, \bibinfo {author} {\bibfnamefont {H.~T.}\ \bibnamefont
  {Stokes}}, \bibinfo {author} {\bibfnamefont {D.~E.}\ \bibnamefont {Tanner}},\
  and\ \bibinfo {author} {\bibfnamefont {D.~M.}\ \bibnamefont {Hatch}},\ }\href
  {http://stokes.byu.edu/iso/isodistort.php} {\bibfield  {journal} {\bibinfo
  {journal} {Journal of Applied Crystallography}\ }\textbf {\bibinfo {volume}
  {39}},\ \bibinfo {pages} {607} (\bibinfo {year} {2006})}\BibitemShut
  {NoStop}%
\bibitem [{\citenamefont {Fecher}\ \emph {et~al.}(2022)\citenamefont {Fecher},
  \citenamefont {K{\"u}bler},\ and\ \citenamefont
  {Felser}}]{Fecher2022chirality}%
  \BibitemOpen
  \bibfield  {author} {\bibinfo {author} {\bibfnamefont {G.~H.}\ \bibnamefont
  {Fecher}}, \bibinfo {author} {\bibfnamefont {J.}~\bibnamefont {K{\"u}bler}},\
  and\ \bibinfo {author} {\bibfnamefont {C.}~\bibnamefont {Felser}},\ }\href
  {https://doi.org/10.3390/ma15175812} {\bibfield  {journal} {\bibinfo
  {journal} {Materials}\ }\textbf {\bibinfo {volume} {15}},\ \bibinfo {pages}
  {5812} (\bibinfo {year} {2022})}\BibitemShut {NoStop}%
\bibitem [{\citenamefont {A.~Jha}\ and\ \citenamefont
  {Joshi}(2012)}]{jhajose2012}%
  \BibitemOpen
  \bibfield  {author} {\bibinfo {author} {\bibfnamefont {G.~J. T. T. F. C. J.
  H. J.~L.}\ \bibnamefont {A.~Jha}, \bibfnamefont {B.~D. O.~Richards}}\ and\
  \bibinfo {author} {\bibfnamefont {P.}~\bibnamefont {Joshi}},\ }\href
  {https://doi.org/10.1179/1743280412Y.0000000005} {\bibfield  {journal}
  {\bibinfo  {journal} {Int. Mater. Rev.}\ }\textbf {\bibinfo {volume} {57}},\
  \bibinfo {pages} {357} (\bibinfo {year} {2012})}\BibitemShut {NoStop}%
\bibitem [{\citenamefont {Weck}\ and\ \citenamefont
  {Kim}(2024)}]{philipkim2024}%
  \BibitemOpen
  \bibfield  {author} {\bibinfo {author} {\bibfnamefont {P.~F.}\ \bibnamefont
  {Weck}}\ and\ \bibinfo {author} {\bibfnamefont {E.}~\bibnamefont {Kim}},\
  }\href {https://doi.org/10.1063/5.0235106} {\bibfield  {journal} {\bibinfo
  {journal} {J. Appl. Phys.}\ }\textbf {\bibinfo {volume} {136}},\ \bibinfo
  {pages} {185901} (\bibinfo {year} {2024})}\BibitemShut {NoStop}%
\bibitem [{\citenamefont {Kondratyuk}\ \emph {et~al.}(1987)\citenamefont
  {Kondratyuk}, \citenamefont {Muradyan}, \citenamefont {Pisarevskii},\ and\
  \citenamefont {Simonov}}]{kondratyuk1987}%
  \BibitemOpen
  \bibfield  {author} {\bibinfo {author} {\bibfnamefont {I.~P.}\ \bibnamefont
  {Kondratyuk}}, \bibinfo {author} {\bibfnamefont {L.~A.}\ \bibnamefont
  {Muradyan}}, \bibinfo {author} {\bibfnamefont {I.~V.}\ \bibnamefont
  {Pisarevskii}},\ and\ \bibinfo {author} {\bibfnamefont {V.~I.}\ \bibnamefont
  {Simonov}},\ }\href@noop {} {\bibfield  {journal} {\bibinfo  {journal}
  {Kristallogr.}\ }\textbf {\bibinfo {volume} {32}},\ \bibinfo {pages} {609}
  (\bibinfo {year} {1987})}\BibitemShut {NoStop}%
\bibitem [{\citenamefont {A.P.~Mirgorodsky}\ and\ \citenamefont
  {Frit}(2000)}]{Mirgorodsky2000}%
  \BibitemOpen
  \bibfield  {author} {\bibinfo {author} {\bibfnamefont {J.-C. C. P.~T.}\
  \bibnamefont {A.P.~Mirgorodsky}, \bibfnamefont {T.~Merle-Méjean}}\ and\
  \bibinfo {author} {\bibfnamefont {B.}~\bibnamefont {Frit}},\ }\href
  {https://doi.org/10.1016/S0022-3697(99)00263-2} {\bibfield  {journal}
  {\bibinfo  {journal} {Journal of Physics and Chemistry of Solids}\ }\textbf
  {\bibinfo {volume} {61}},\ \bibinfo {pages} {501} (\bibinfo {year}
  {2000})}\BibitemShut {NoStop}%
\bibitem [{\citenamefont {Champarnaud-Mesjard}\ \emph
  {et~al.}(2000)\citenamefont {Champarnaud-Mesjard}, \citenamefont
  {Blanchandin}, \citenamefont {Thomas}, \citenamefont {Mirgorodsky},
  \citenamefont {Merle-M{\'e}jean},\ and\ \citenamefont
  {Frit}}]{champarnaud2000}%
  \BibitemOpen
  \bibfield  {author} {\bibinfo {author} {\bibfnamefont {J.~C.}\ \bibnamefont
  {Champarnaud-Mesjard}}, \bibinfo {author} {\bibfnamefont {S.}~\bibnamefont
  {Blanchandin}}, \bibinfo {author} {\bibfnamefont {P.}~\bibnamefont {Thomas}},
  \bibinfo {author} {\bibfnamefont {A.}~\bibnamefont {Mirgorodsky}}, \bibinfo
  {author} {\bibfnamefont {T.}~\bibnamefont {Merle-M{\'e}jean}},\ and\ \bibinfo
  {author} {\bibfnamefont {B.}~\bibnamefont {Frit}},\ }\href
  {https://doi.org/10.1016/S0022-3697(00)00012-3} {\bibfield  {journal}
  {\bibinfo  {journal} {J. Phys. Chem. Solids}\ }\textbf {\bibinfo {volume}
  {61}},\ \bibinfo {pages} {1499} (\bibinfo {year} {2000})}\BibitemShut
  {NoStop}%
\bibitem [{\citenamefont {Yanlu~Li}\ and\ \citenamefont
  {Zhao}(2010)}]{yfancheng2010}%
  \BibitemOpen
  \bibfield  {author} {\bibinfo {author} {\bibfnamefont {H.~S.-X. C. P.~L.}\
  \bibnamefont {Yanlu~Li}, \bibfnamefont {Weiliu~Fan}}\ and\ \bibinfo {author}
  {\bibfnamefont {X.}~\bibnamefont {Zhao}},\ }\href
  {https://doi.org/10.1063/1.3406135} {\bibfield  {journal} {\bibinfo
  {journal} {J. Appl. Phys.}\ }\textbf {\bibinfo {volume} {107}},\ \bibinfo
  {pages} {093506} (\bibinfo {year} {2010})}\BibitemShut {NoStop}%
\bibitem [{\citenamefont {J.C.~Sabadel}\ and\ \citenamefont
  {Philippot}(1997)}]{Sabadel1997}%
  \BibitemOpen
  \bibfield  {author} {\bibinfo {author} {\bibfnamefont {D.~C.-H. P. B. O. D.
  A.~I.}\ \bibnamefont {J.C.~Sabadel}, \bibfnamefont {P.~Armand}}\ and\
  \bibinfo {author} {\bibfnamefont {E.}~\bibnamefont {Philippot}},\ }\href
  {https://doi.org/https://doi.org/10.1006/jssc.1997.7499} {\bibfield
  {journal} {\bibinfo  {journal} {J. Solid State Chem.}\ }\textbf {\bibinfo
  {volume} {132}},\ \bibinfo {pages} {411} (\bibinfo {year}
  {1997})}\BibitemShut {NoStop}%
\bibitem [{\citenamefont {Volker L.~Deringer}\ and\ \citenamefont
  {Dronskowski}(2014)}]{Deringer2014}%
  \BibitemOpen
  \bibfield  {author} {\bibinfo {author} {\bibfnamefont {R.~P.~S.}\
  \bibnamefont {Volker L.~Deringer}}\ and\ \bibinfo {author} {\bibfnamefont
  {R.}~\bibnamefont {Dronskowski}},\ }\href {https://doi.org/10.1021/cg401822g}
  {\bibfield  {journal} {\bibinfo  {journal} {Cryst. Growth Des.}\ }\textbf
  {\bibinfo {volume} {14}},\ \bibinfo {pages} {871} (\bibinfo {year}
  {2014})}\BibitemShut {NoStop}%
\bibitem [{\citenamefont {Beyer}(1967)}]{Beyer1967}%
  \BibitemOpen
  \bibfield  {author} {\bibinfo {author} {\bibfnamefont {V.~H.}\ \bibnamefont
  {Beyer}},\ }\href {https://doi.org/10.1524/zkri.1967.124.3.228} {\bibfield
  {journal} {\bibinfo  {journal} {Kristallogr.}\ }\textbf {\bibinfo {volume}
  {124}},\ \bibinfo {pages} {228} (\bibinfo {year} {1967})}\BibitemShut
  {NoStop}%
\bibitem [{\citenamefont {Peercy}\ and\ \citenamefont
  {Fritz}(1974)}]{Peercy1974}%
  \BibitemOpen
  \bibfield  {author} {\bibinfo {author} {\bibfnamefont {P.~S.}\ \bibnamefont
  {Peercy}}\ and\ \bibinfo {author} {\bibfnamefont {I.~J.}\ \bibnamefont
  {Fritz}},\ }\href {https://doi.org/10.1103/PhysRevLett.32.466} {\bibfield
  {journal} {\bibinfo  {journal} {Phys. Rev. Lett.}\ }\textbf {\bibinfo
  {volume} {32}},\ \bibinfo {pages} {466} (\bibinfo {year} {1974})}\BibitemShut
  {NoStop}%
\bibitem [{\citenamefont {Worlton}\ and\ \citenamefont
  {Beyerlein}(1975)}]{Worlton1975}%
  \BibitemOpen
  \bibfield  {author} {\bibinfo {author} {\bibfnamefont {T.~G.}\ \bibnamefont
  {Worlton}}\ and\ \bibinfo {author} {\bibfnamefont {R.~A.}\ \bibnamefont
  {Beyerlein}},\ }\href {https://doi.org/10.1103/PhysRevB.12.1899} {\bibfield
  {journal} {\bibinfo  {journal} {Phys. Rev. B}\ }\textbf {\bibinfo {volume}
  {12}},\ \bibinfo {pages} {1899} (\bibinfo {year} {1975})}\BibitemShut
  {NoStop}%
\bibitem [{\citenamefont {Skelton}\ \emph {et~al.}(1976)\citenamefont
  {Skelton}, \citenamefont {Feldman}, \citenamefont {Liu},\ and\ \citenamefont
  {Spain}}]{Skelton1976}%
  \BibitemOpen
  \bibfield  {author} {\bibinfo {author} {\bibfnamefont {E.~F.}\ \bibnamefont
  {Skelton}}, \bibinfo {author} {\bibfnamefont {J.~L.}\ \bibnamefont
  {Feldman}}, \bibinfo {author} {\bibfnamefont {C.~Y.}\ \bibnamefont {Liu}},\
  and\ \bibinfo {author} {\bibfnamefont {I.~L.}\ \bibnamefont {Spain}},\ }\href
  {https://doi.org/10.1103/PhysRevB.13.2605} {\bibfield  {journal} {\bibinfo
  {journal} {Phys. Rev. B}\ }\textbf {\bibinfo {volume} {13}},\ \bibinfo
  {pages} {2605} (\bibinfo {year} {1976})}\BibitemShut {NoStop}%
\bibitem [{\citenamefont {Jayaraman}\ and\ \citenamefont
  {Kourouklis}(1991)}]{Jayaraman1991}%
  \BibitemOpen
  \bibfield  {author} {\bibinfo {author} {\bibfnamefont {A.}~\bibnamefont
  {Jayaraman}}\ and\ \bibinfo {author} {\bibfnamefont {G.~A.}\ \bibnamefont
  {Kourouklis}},\ }\href {https://doi.org/10.1007/BF02845698} {\bibfield
  {journal} {\bibinfo  {journal} {Pramana - J. Phys.}\ }\textbf {\bibinfo
  {volume} {36}},\ \bibinfo {pages} {133} (\bibinfo {year} {1991})}\BibitemShut
  {NoStop}%
\bibitem [{\citenamefont {Majzlan}\ \emph {et~al.}(2022)\citenamefont
  {Majzlan}, \citenamefont {Notz}, \citenamefont {Haase}, \citenamefont
  {Kamitsos}, \citenamefont {Tagiara},\ and\ \citenamefont
  {Dachs}}]{Majzlan2022}%
  \BibitemOpen
  \bibfield  {author} {\bibinfo {author} {\bibfnamefont {J.}~\bibnamefont
  {Majzlan}}, \bibinfo {author} {\bibfnamefont {S.}~\bibnamefont {Notz}},
  \bibinfo {author} {\bibfnamefont {P.}~\bibnamefont {Haase}}, \bibinfo
  {author} {\bibfnamefont {E.~I.}\ \bibnamefont {Kamitsos}}, \bibinfo {author}
  {\bibfnamefont {N.~S.}\ \bibnamefont {Tagiara}},\ and\ \bibinfo {author}
  {\bibfnamefont {E.}~\bibnamefont {Dachs}},\ }\href
  {https://doi.org/10.1016/j.chemer.2022.125915} {\bibfield  {journal}
  {\bibinfo  {journal} {Geochemistry}\ }\textbf {\bibinfo {volume} {82}},\
  \bibinfo {pages} {125915} (\bibinfo {year} {2022})}\BibitemShut {NoStop}%
\bibitem [{\citenamefont {Ceriotti}\ \emph {et~al.}(2006)\citenamefont
  {Ceriotti}, \citenamefont {Pietrucci},\ and\ \citenamefont
  {Bernasconi}}]{Ceriotti2006teo2}%
  \BibitemOpen
  \bibfield  {author} {\bibinfo {author} {\bibfnamefont {M.}~\bibnamefont
  {Ceriotti}}, \bibinfo {author} {\bibfnamefont {F.}~\bibnamefont
  {Pietrucci}},\ and\ \bibinfo {author} {\bibfnamefont {M.}~\bibnamefont
  {Bernasconi}},\ }\href {https://doi.org/10.1103/PhysRevB.73.104304}
  {\bibfield  {journal} {\bibinfo  {journal} {Phys. Rev. B}\ }\textbf {\bibinfo
  {volume} {73}},\ \bibinfo {pages} {104304} (\bibinfo {year}
  {2006})}\BibitemShut {NoStop}%
\bibitem [{\citenamefont {Walsh}\ \emph {et~al.}(2011)\citenamefont {Walsh},
  \citenamefont {Payne}, \citenamefont {Egdell},\ and\ \citenamefont
  {Watson}}]{Walsh2011lonepair}%
  \BibitemOpen
  \bibfield  {author} {\bibinfo {author} {\bibfnamefont {A.}~\bibnamefont
  {Walsh}}, \bibinfo {author} {\bibfnamefont {D.~J.}\ \bibnamefont {Payne}},
  \bibinfo {author} {\bibfnamefont {R.~G.}\ \bibnamefont {Egdell}},\ and\
  \bibinfo {author} {\bibfnamefont {G.~W.}\ \bibnamefont {Watson}},\ }\href
  {https://doi.org/10.1039/C1CS15098G} {\bibfield  {journal} {\bibinfo
  {journal} {Chemical Society Reviews}\ }\textbf {\bibinfo {volume} {40}},\
  \bibinfo {pages} {4455} (\bibinfo {year} {2011})}\BibitemShut {NoStop}%
\bibitem [{\citenamefont {Bersuker}(2021)}]{Bersuker2021pjte}%
  \BibitemOpen
  \bibfield  {author} {\bibinfo {author} {\bibfnamefont {I.~B.}\ \bibnamefont
  {Bersuker}},\ }\href {https://doi.org/10.1021/acs.chemrev.0c00718} {\bibfield
   {journal} {\bibinfo  {journal} {Chemical Reviews}\ }\textbf {\bibinfo
  {volume} {121}},\ \bibinfo {pages} {1463} (\bibinfo {year} {2021})},\
  \bibinfo {note} {published online December 22, 2020}\BibitemShut {NoStop}%
\bibitem [{\citenamefont {Bersuker}(2013)}]{Bersuker2013formula}%
  \BibitemOpen
  \bibfield  {author} {\bibinfo {author} {\bibfnamefont {I.~B.}\ \bibnamefont
  {Bersuker}},\ }\href {https://doi.org/10.1021/cr300279n} {\bibfield
  {journal} {\bibinfo  {journal} {Chemical Reviews}\ }\textbf {\bibinfo
  {volume} {113}},\ \bibinfo {pages} {1351} (\bibinfo {year}
  {2013})}\BibitemShut {NoStop}%
\bibitem [{cec()}]{ceci}%
  \BibitemOpen
  \href {http://www.ceci-hpc.be/} {\bibinfo {title} {Ceci high performance
  computing}}\BibitemShut {NoStop}%
\end{thebibliography}
%apsrev4-2.bst 2019-01-14 (MD) hand-edited version of apsrev4-1.bst
%Control: key (0)
%Control: author (72) initials jnrlst
%Control: editor formatted (1) identically to author
%Control: production of article title (-1) disabled
%Control: page (0) single
%Control: year (1) truncated
%Control: production of eprint (0) enabled
%

\end{document}